\documentclass[iop]{emulateapj}
\usepackage{float}
\usepackage{color}
\usepackage{booktabs}
\usepackage{amsmath}
\newcommand{\ra}[1]{\renewcommand{\arraystretch}{#1}}

\def \sigfifteen {\hbox{$\sigma_{\rm DM,15}$}}
\def \mtwoh {\hbox{$M_{200c}$}}
\def \sigdm {\hbox{$\sigma_{\rm DM}$}}

\shortauthors{Wang et al.}
\shorttitle{A galaxy cluster at $z=2.506$}

\newcommand{\Msol}{\hbox{$M_\odot$}}

\begin{document} 

\title{Discovery of a galaxy cluster with a violently starbursting core at $\MakeLowercase{z}=2.506$}

\author{
Tao~Wang$\!$\altaffilmark{1,2}, 
David~Elbaz$\!$\altaffilmark{1}, 
Emanuele~Daddi$\!$\altaffilmark{1}, 
Alexis~Finoguenov$\!$\altaffilmark{3}, 
Daizhong~Liu$\!$\altaffilmark{4}, 
Corentin~Schreiber$\!$\altaffilmark{5}, 
Sergio~Mart\'in$\!$\altaffilmark{6,7,8}, 
Veronica~Strazzullo$\!$\altaffilmark{9}, 
Francesco~Valentino$\!$\altaffilmark{1}, 
Remco~van der Burg$\!$\altaffilmark{1}, 
Anita~Zanella$\!$\altaffilmark{1}, 
Laure~Ciesla$\!$\altaffilmark{1}, 
Raphael~Gobat$\!$\altaffilmark{10}, 
Amandine~Le Brun$\!$\altaffilmark{1}, 
Maurilio~Pannella$\!$\altaffilmark{9}, 
Mark~Sargent$\!$\altaffilmark{11}, 
Xinwen~Shu$\!$\altaffilmark{12}, 
Qinghua~Tan$\!$\altaffilmark{4},
Nico Cappelluti$\!$\altaffilmark{13},
Yanxia Li$\!$\altaffilmark{14}
}

\altaffiltext{1}{Laboratoire AIM-Paris-Saclay, CEA/DSM/Irfu, CNRS, Universit\'e Paris Diderot, Saclay, pt courrier 131, 91191 Gif-sur-Yvette, France}
\altaffiltext{2}{Key Laboratory of Modern Astronomy and Astrophysics in Ministry of Education, School of Astronomy and Space Sciences, Nanjing University, Nanjing, 210093, China }
\altaffiltext{3}{Department of Physics, University of Helsinki, Gustaf H{\"a}llstr{\"o}min katu 2a, FI-0014 Helsinki, Finland}
\altaffiltext{4}{Purple Mountain Observatory, Chinese Academy of Sciences, 2 West Beijing Road, Nanjing 210008, China}
\altaffiltext{5}{Leiden Observatory, Leiden University, NL-2300 RA Leiden, The Netherlands}
\altaffiltext{6}{European Southern Observatory, Alonso de C{\'o}rdova 3107, Vitacura, Santiago, Chile}
\altaffiltext{7}{Joint ALMA Observatory, Alonso de C{\'o}rdova 3107, Vitacura, Santiago, Chile}
\altaffiltext{8}{Institut de Radio Astronomie Millim{\'e}trique, 300 rue de la Piscine, Dom. Univ., F-38406, St. Martin d'H{\`e}res, France}
\altaffiltext{9}{Department of Physics, Ludwig-Maximilians-Universit{\"a}t, Scheinerstr. 1, D-81679 M{\"u}nchen, Germany}
\altaffiltext{10}{School of Physics, Korea Institute for Advanced Study, Hoegiro 85, Dongdaemun-gu, Seoul 130-722, Republic of Korea}
\altaffiltext{11}{Department of Physics and Astronomy, University of Sussex, Brighton BN1 9QH, UK}
\altaffiltext{12}{Department of Physics, Anhui Normal University, Wuhu, Anhui, 241000, China}
\altaffiltext{13}{Department of Astronomy, Yale University, New Haven, CT 06511, USA}
\altaffiltext{14}{Institute for Astronomy, University of Hawaii, 2680 Woodlawn Drive, Honolulu, HI 96822, USA}
\email{tao.wang@cea.fr}

\begin{abstract}

We report the discovery of a remarkable concentration of massive galaxies with extended X-ray emission at $z_{spec} = 2.506$, which contains 11 massive ($M_{*} \gtrsim 10^{11} M_{\odot}$) galaxies in the central 80kpc region (11.6$\sigma$ overdensity). We have spectroscopically confirmed 17 member galaxies with 11 from CO and the remaining ones from $H\alpha$. The X-ray luminosity, stellar mass content and velocity dispersion all point to a collapsed, cluster-sized dark matter halo with mass $M_{200c} = 10^{13.9\pm0.2} M_{\odot}$, making it the most distant X-ray-detected cluster known to date. Unlike other clusters discovered so far, this structure is dominated by star-forming galaxies (SFGs) in the core with only 2 out of the 11 massive galaxies classified as quiescent. The star formation rate (SFR) in the 80kpc core reaches $\sim$3400 $M_{\odot}$ yr$^{-1}$ with a gas depletion time of $\sim 200$ Myr, suggesting that we caught this cluster in rapid build-up of a dense core. The high SFR is driven by both a high abundance of SFGs and a higher starburst fraction ($\sim25\%$, compared to 3\%-5\% in the field). The presence of both a collapsed, cluster-sized halo and a predominant population of massive SFGs suggests that this structure could represent an important transition phase between protoclusters and mature clusters. It provides evidence that the main phase of massive galaxy passivization will take place after galaxies accrete onto the cluster, providing new insights into massive cluster formation at early epochs. The large integrated stellar mass at such high redshift challenges our understanding of massive cluster formation. 

\end{abstract}
\clearpage
\section{Introduction} 

Clusters of galaxies represent the densest environments and trace the most massive dark matter (DM) halos in the universe. Studying the formation and evolution of galaxy clusters and their member galaxies is fundamental to our understanding of both galaxy formation and cosmology~\citep{Kravtsov:2012}. 
Massive galaxy clusters in the local universe are characterized by a significant population of massive, passive ellipticals in their cores. Galaxy cluster archeology and numerical simulations suggest that these massive clusters and their member galaxies have experienced a rapid formation phase at $z > 2$, when the bulk of the stars in central cluster galaxies was formed~\citep{Thomas:2005,DeLucia:2007}, and the first collapsed, cluster-sized halos (progenitors of today's most massive galaxy clusters) with masses $M_{200c} \gtrsim 10^{14} M_{\odot}$ were assembled~\citep{Chiang:2013}. Observations of galaxy structures in this rapid formation phase are critical to map the full path of galaxy cluster formation and to answer fundamental questions about the effect of dense environments on galaxy formation and evolution. 
Such structures, however, have been so far difficult to detect due to their rareness and distance.

Extensive efforts have been made to search for high-redshift structures during the last decade with a variety of techniques, and a number of galaxy (proto)clusters at $z \gtrsim 1.5-2$ have been discovered. A few of these structures found up to $z \sim 2$ already show evidence of a collapsed, cluster-sized  halo and exhibit a high concentration of quiescent galaxies in the core (with a well-defined red sequence,~\citealt{Papovich:2010,Gobat:2011,Stanford:2012,Andreon:2014,Newman:2014}), hence they can be classified as bona fide mature clusters. Some of them still contain a substantial number of star-forming galaxies (SFGs)~\citep{Brodwin:2013,Gobat:2013,Strazzullo:2013,Clements:2014,Webb:2015,Valentino:2015}, and a few of them show clear evidence of  a reversal of the star formation-density relation~\citep{Elbaz:2007} with enhanced star formation in cluster members with respect to field galaxies~\citep{Tran:2010,Santos:2015}. However, most of these clusters are already dominated by quiescent galaxies in the core, at least at the massive end ($M_{*} \gtrsim 10^{11} M_{\odot}$), with a significantly higher quiescent fraction compared to the field. Hence to probe the main formation epoch of the most massive cluster galaxies, we need to explore even higher redshifts, i.e., $z > 2$.

Most currently known $z \gtrsim 2$ structures exhibit lower galaxy number densities and are spread in multiple, less massive, and not collapsed halos compared to mature clusters~\citep{Steidel:1998,Kurk:2000,Venemans:2007,Chapman:2009,Miley:2008,Daddi:2009a,Tanaka:2011,Capak:2011,Trenti:2012,Hayashi:2012,Spitler:2012,Diener:2013,Chiang:2013,Koyama:2013,Lemaux:2014,YuanT:2014,Mei:2015,Casey:2015,Kubo:2016}. 
These structures are believed to be in various early evolutionary stages of cluster formation, and are usually called ``protoclusters.'' Although there is no consensus on the distinction between protoclusters and clusters, recent works suggest that the lack of a collapsed, cluster-sized halo ($M_{200c} \gtrsim 10^{14} M_{\odot}$) is a key feature to differentiate protoclusters from clusters~\citep{Diener:2015,Muldrew:2015}. This distinction could be important since the dominating environmental process that shapes galaxy evolution depends on the mass of the host DM halo, such as ram pressure stripping~\citep{Gunn:1972} and strangulation~\citep{Larson:1980}. Moreover, in contrast to mature galaxy clusters, most of the galaxies in protoclusters are found to be star-forming, with no clear evidence of an elevated quiescent fraction compared to field galaxies.   
Hence the transition between protoclusters and mature clusters requires both the collapse of a massive, cluster-sized halo and the formation and quenching of a significant population of massive galaxies. 
Clear evidence for galaxy structures in such a rapid transition phase, however, is lacking from current observations.

From a theoretical perspective, the halo assembly history of today's massive clusters is relatively well understood. Numerical simulations suggest that the progenitor of a present day ``Coma''-type galaxy cluster ($M_{200c} > 10^{15} M_{\odot}$) exhibits overdensities of galaxies over an extended area, $\gtrsim$ 25 Mpc at $z > 2$, and consists of many separated halos~\citep{Chiang:2013,Muldrew:2015,Contini:2016}. Among these halos, the most massive one could reach a few times 10$^{13}$ to $10^{14} M_{\odot}$, and should be detected as a cluster. These results are in good agreement with observations. However, details on the build-up of the stellar mass content of massive clusters and the physical mechanisms leading to the distinct galaxy population in clusters and the field are still under active debate. Much of these debates focus on the relative importance of different environmental effects on massive galaxy evolution. For instance, it remains unclear whether the bulk population of central cluster ellipticals are formed after galaxies have become part of a cluster (due to, e.g., ram pressure stripping and frequent mergers), or whether they are already established due to ``pre-processing'' in grouplike environment before their accretion onto a cluster-sized halo~(due to, e.g., strangulation). A few recent theoretical studies provide some insights into this issue, however, reaching different conclusions~\citep{Balogh:2009,Berrier:2009,McGee:2009,DeLucia:2012,Granato:2015,Contini:2016}. The major difficulty in constraining these theoretical models comes form the lack of comprehensive understanding of star formation and quenching (quiescent fraction) in halos with different masses at high redshifts. The fact that most galaxy clusters up to $z \sim 2$ are already dominated by quiescent galaxies in the core suggests that we need to explore structures at even higher redshift to put observational constraints on this issue.

In this paper we report the discovery of CL J1001+0220~(CL J1001, hereafter), a remarkable concentration of massive SFGs at $z=2.506$ with 17 spectroscopic members. The detection of extended X-ray emission and the velocity dispersion of its member galaxies are suggestive of a virialized, cluster-sized halo with $M_{200c} \sim  10^{13.9} M_{\odot}$, making it the most distant X-ray-detected galaxy cluster known to date. However, unlike any clusters detected so far, the core of this structure is dominated by SFGs with a star formation rate (SFR) density of $\sim 3400 M_{\odot}$ yr$^{-1}$ in the central 80 kpc region, suggesting that most of the ellipticals in this cluster will form after galaxies accrete onto the cluster core. This provides one of the first observational constraints on the role of ``pre-processing'' in the early formation of massive clusters.

This paper is organized as follows. We describe the target selection and multiwavelength imaging of CL J1001 in Section~\ref{Sec:detection}. Spectroscopic follow-up observations and redshift determinations are shown in Section~\ref{Sec:spectra}. We present the X-ray observations of the cluster from $Chandra$ and $XMM-Newton$ in Section~\ref{Sec:Xray}. In Section~\ref{Sec:cluster}, we discuss the global properties of the cluster. In Section~\ref{Sec:galaxies}, we explore physical properties of its member galaxies. We then discuss the implications of this cluster on galaxy and cluster formation, as well as on cosmology, in Section~\ref{Sec:discussion}.  Section~\ref{Sec:conclusion} summarizes our main results. Unless specified otherwise, all magnitudes are in the AB system, and we assume cosmological parameters of $H_{0}$ = 70 km s$^{-1}$ Mpc$^{-1}$, $\Omega_{M}$ = 0.3, and $\Omega_{\Lambda}$ = 0.7. A \cite{Salpeter:1955} initial mass function (IMF) is adopted to derive stellar masses and SFRs. When necessary, we converted literature values of stellar masses and SFRs based on the \cite{Chabrier:2003}  IMF to Salpeter by multiplying by a factor of 1.74 (0.24 dex). Throughout this paper, we define $M_{200c}$ as the total halo mass contained within $R_{200c}$,  the radius from the cluster center within which the average density is 200 times the critical density at the cluster redshift. 

\section{Target selection and multwavelength imaging}
\label{Sec:detection}
\begin{figure*}[!tbh]
\centering
\includegraphics[scale=0.49]{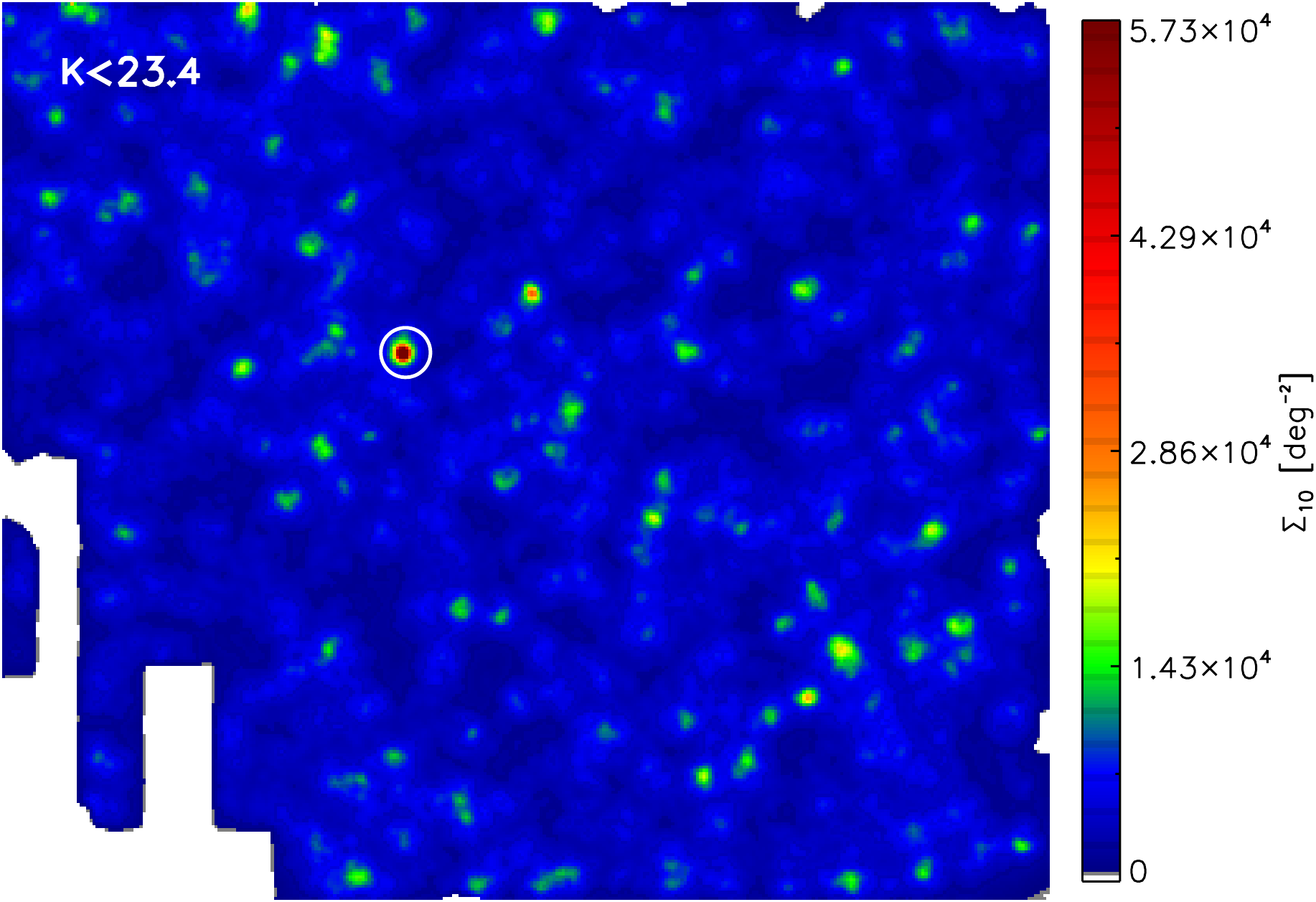}
\includegraphics[scale=0.49]{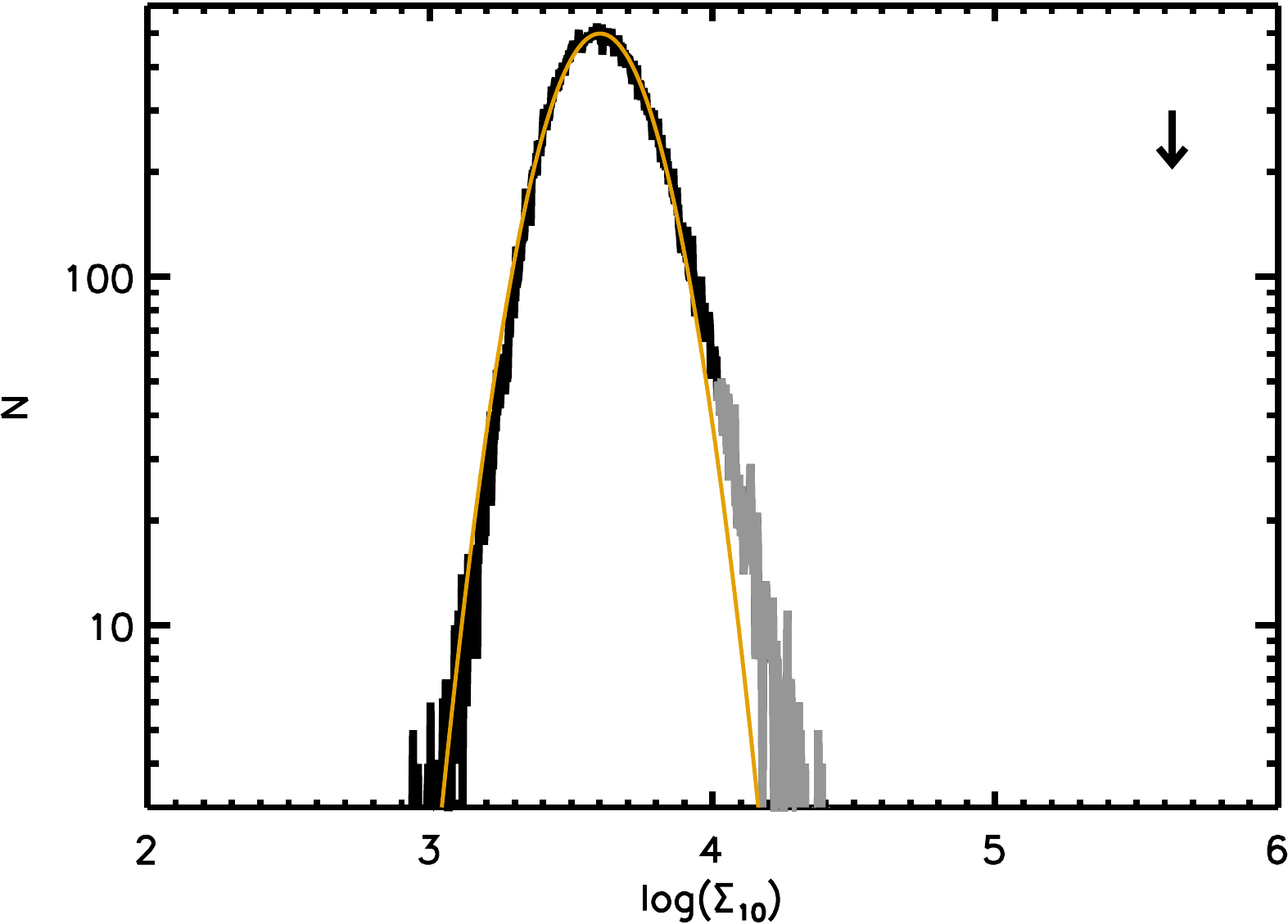}
\caption{
\textbf{Left}: A smoothed map of $\Sigma_{10}$ for the whole sample of DRGs in COSMOS. The parts of the map that are affected by image borders or bad areas are masked out. The white circle with a radius of 2$^{\prime}$ denotes the overdensity region studied in this paper, which has the highest $\Sigma_{10}$ value. \textbf{Right}: the distribution of log~$\Sigma_{10}$ values in the map and its Gaussian fit (white and orange lines, respectively). The grayed-out part of the distribution is not considered in the fit to avoid overdensities  affecting the fitting results. The white arrow shows the peak $\Sigma_{10}$ value of the overdensity.
\label{Fig:significance}}
\end{figure*}

Searching for overdensities of massive galaxies represents a relatively unbiased way of identifying galaxy (proto)clusters at high redshifts. A number of (proto)clusters have been identified based on overdensities of color-selected massive galaxies~\citep{Papovich:2010}. In particular, the distant red galaxy (DRGs, \cite{Franx:2003}) population has been shown to be quite efficient at selecting massive galaxies (including both quiescent and star-forming ones) at $z > 2$~\citep{vanDokkum:2003}. To search for galaxy clusters at $z > 2$, 
We have performed a systematic study of overdensities of DRGs with $J - K_{s} > 1.3$ in the COSMOS field using a $K_{s}$-band selected catalog~($K_{s} < 23.4$, 90\% completeness,~\citealt{McCracken:2012,Muzzin:2013a}). Overdensities of DRGs have been shown to be good tracers of potential massive structures at $z > 2$~\citep{Uchimoto:2012}. 
We constructed a galaxy surface density map using a local galaxy density indicator, $\Sigma_{10}$, which is $10/(\pi r_{10}^{2})$ with $r_{10}$ the distance to the 10th nearest neighbor on a grid (Fig~\ref{Fig:significance}). We then fitted the distribution of log($\Sigma_{10}$) values in the map with a  gaussian function, as shown in the right panel of Fig~\ref{Fig:significance}. The most significant outlier (centered at R.A.=10:00:57.13, decl. = +02:20:11.83) of the best-fit gaussian distribution exhibits a $\Sigma_{10} \sim 11.6 \sigma$ higher than the mean. We have also tried to use different density estimators, e.g., $\Sigma_{5}$, which is $5/(\pi r_{5}^{2})$ with $r_{5}$ the distance to the 5th nearest neighbor, yielding similar significance for this overdensity. This overdensity includes 11 DRGs and 2 blue galaxies within a 10" radius, or 80 kpc at $z = 2.5$. The photometric redshift distribution of these 13 galaxies shows a prominent peak at $z \sim 2.5$ with one of them identified as a Lyman-$\alpha$ emitter at $z \sim 2.5\pm0.1$ based on intermediate-band data (IA427 filter) in the Subaru COSMOS 20 survey~\citep{Taniguchi:2015}.

\begin{figure}[!tbh]
\centering
\includegraphics[scale=0.4]{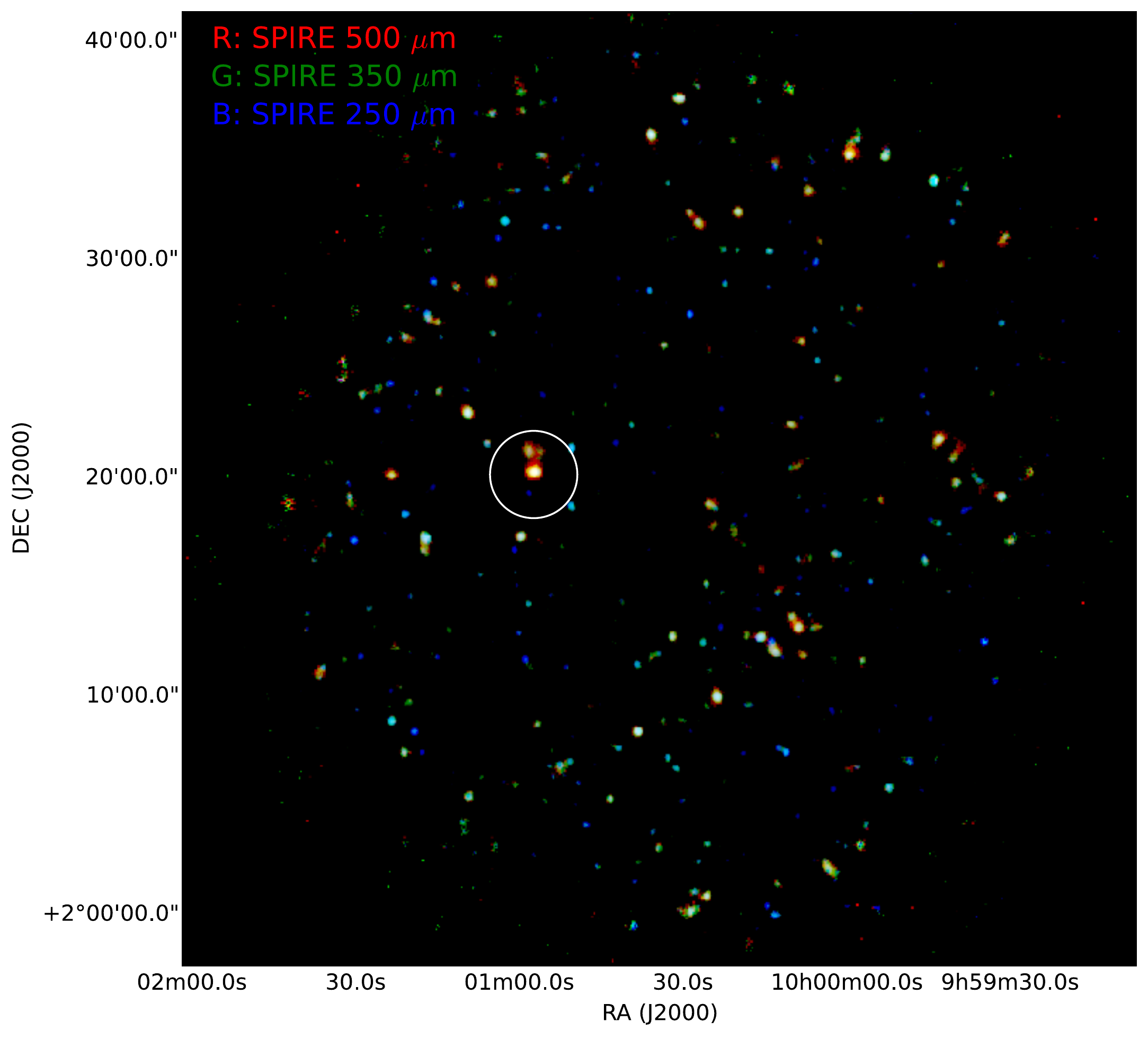}
\caption{RGB $Herschel$/SPIRE composite color image of the COSMOS field covered by the CANDELS-$Herschel$ survey. The  $R$, $G$ and $B$ channels correspond to SPIRE 500, 350 and 250 $\mu$m, respectively. Only sources brighter than 20 mJy at one or more of the three SPIRE wavelengths are shown. Sources with redder colors tend to be at higher redshifts. The large white circle with a radius of 2$\arcmin$ indicates the position of the galaxy overdensity, which is the brightest SPIRE source in the whole field.
\label{Fig:Herschel}}
\end{figure}

\begin{figure*}
\centering
\includegraphics[scale=0.7]{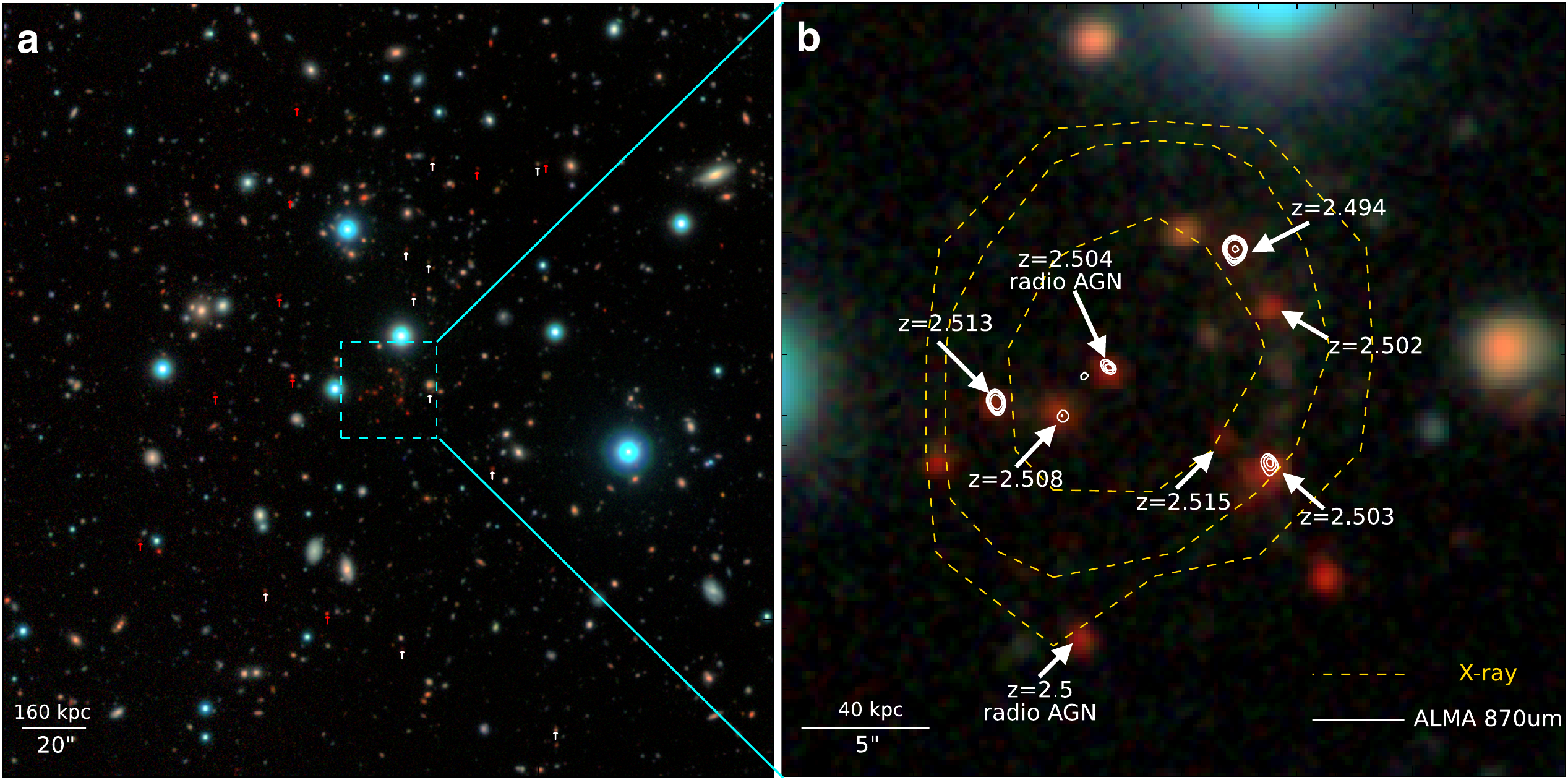}
\caption{RGB composite color image of the region around the cluster core. The  $R$, $G$ and $B$ channel correspond to the $K_{s}$, $J$ and $Y$ bands from the UltraVISTA survey, respectively. The left panel (\textbf{a}) corresponds to a  $4' \times 4'$ region while the right panel (\textbf{b}) is an enlarged image of the central $30^{\prime\prime} \times 30^{\prime\prime}$ region around the cluster core. Red arrows indicate distant red galaxies (DRGs) outside the core with $z_{phot}=2.5\pm0.5$ while white arrows indicate spectroscopically confirmed members within 3$\sigma$ of the peak of the redshift distribution ($z_{spec}=2.506\pm0.018$), including 7 galaxies in the core (indicated in the right panel) and 10 galaxies in the outskirts. Extended X-ray emission (0.5-2 keV)  and ALMA 870 $\mu$m continuum are overlaid, respectively, with yellow and white contours in the right panel. There are 11 DRGs (5 detected with ALMA at 870 $\mu$m) and 2 blue galaxies within the central 10$^{\prime\prime}$ region, or 80 kpc at $z = 2.5$.
\label{Fig:RGB}}
\end{figure*}

The same overdensity also corresponds to the brightest $Herschel$/SPIRE source (unresolved) in the region covered by the CANDELS-$Herschel$ survey (PI: Mark Dickinson) in the COSMOS field (Fig.~\ref{Fig:Herschel}), with flux densities of $\sim$ 61, 77, and 66 mJy at SPIRE 250, 350, and 500 $\mu$m, respectively. With a peak at 350 $\mu$m, the far-infrared spectral energy distribution (SED) of this overdensity provides further evidence that most of its member galaxies are likely at $z \sim 2.5$. This overdensity was also detected at 850 $\mu$m with SCUBA-2~\citep{Casey:2013} and 1.1mm with Aztec~\citep{Aretxaga:2011} with flux densities 14.8 and 8.9 mJy, respectively. 
The same region was also observed as a candidate of lensed sources with ALMA at band-7 (870 $\mu$m) as described in \cite{Bussmann:2015}.  ALMA resolves 5 out of the 11 DRGs in the core down to $S_{870~\mu m} > 1.6$~mJy~(Fig~\ref{Fig:RGB}). These observations suggest that vigorous star formation is taking place in the member galaxies of this structure.  Motivated by its high far-infrared and millimeter flux densities,  we have performed a series of follow-up observations from near-infrared to millimeter to explore properties of this overdensity and its member galaxies.

\section{Spectroscopic observations and redshift determination}
\label{Sec:spectra}

\subsection{IRAM-NOEMA}
We first conducted observations with IRAM-NOEMA to resolve the millimeter emission and measure the redshift of this overdensity through the CO(5-4) line. 
Observations were carried out as part of a DDT program at NOEMA between 2014 November 7th, 2014 and 2015 March 5th.
Aiming at the detection of CO(5-4), we performed a frequency scan between 161.1 and 171.5~GHz, which corresponds to a redshift range of $2.36 < z < 2.58$ sampled by the CO(5-4) line.
We used the 3.6~GHz instantaneous bandwidth of the WideX correlator in three frequency setups centered at 162.9, 166.1, and 169.7~GHz. Between 1 and 1.5 hours were devoted to each setup, reaching an rms of $\sim13$~mJy at the original resolution, or $\sim0.3-0.4$~mJy over an integrated 1000~km~s$^{-1}$ band.
Following up a preliminary detection, we further integrated 1.8 extra hours at 164.45~GHz, and 2.4 hr at 98.68~GHz to detect the CO(3-2) transition.
Gain calibration was performed on the nearby quasars 1055+018 and 0906+015, which were also used for regular pointing and focus measurements.
Calibration was carried out with GILDAS \footnote{http://www.iram.fr/IRAMFR/GILDAS}. 

\begin{figure*}
\centering
\includegraphics[scale=0.7]{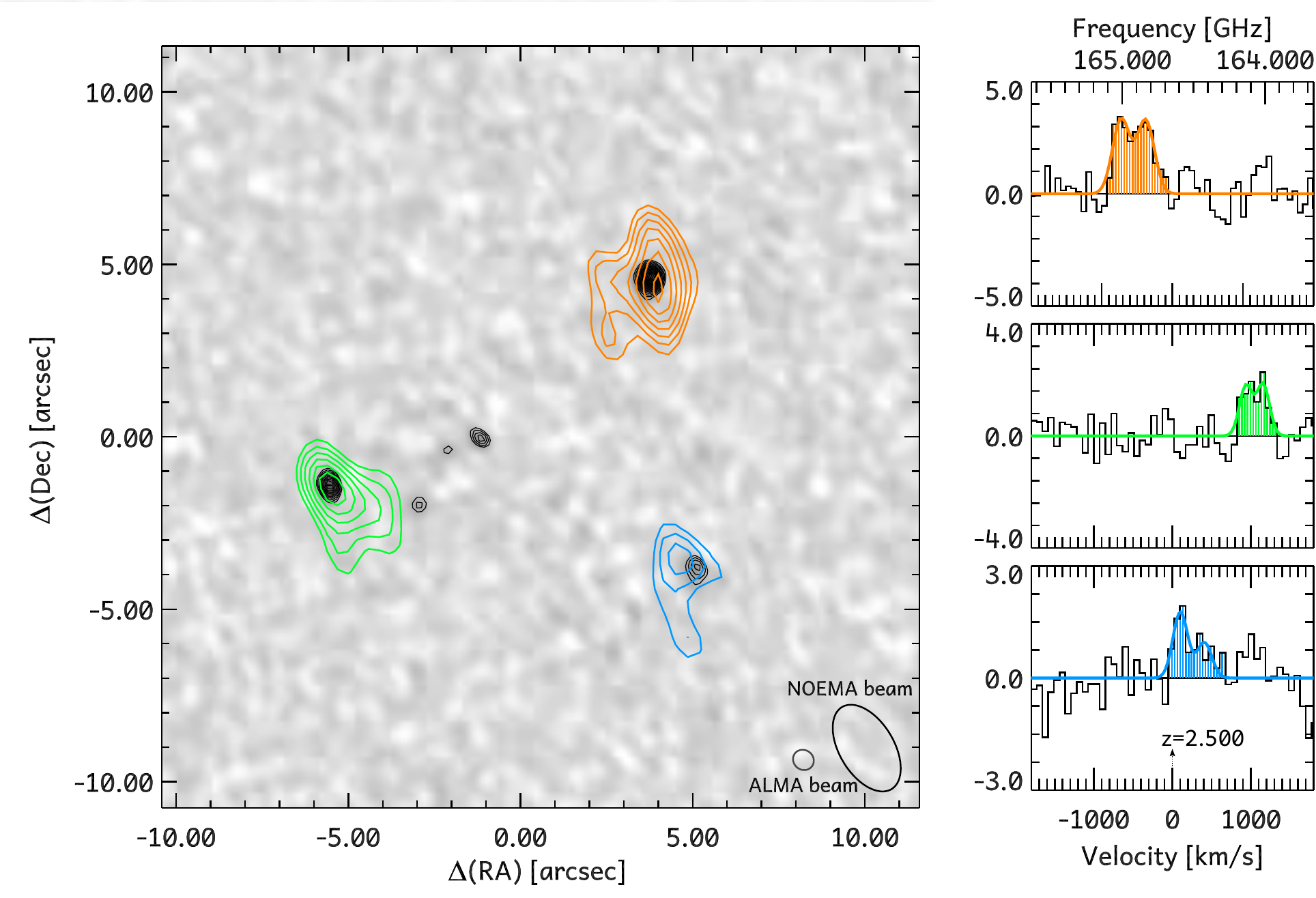}
\caption{ALMA and IRAM-NOEMA observations of the cluster core. ALMA 870 $\mu$m continuum map of the cluster core ($20^{\prime\prime} \times 20^{\prime\prime}$), overlaid with CO(5-4) emission line detections from IRAM-NOEMA. In total, five sources in the cluster are detected with ALMA, three of which are also detected in CO(5-4) with NOEMA. The ALMA and NOEMA beams are denoted by the small and large ellipses, respectively. The CO(5-4) line spectra for the three sources are shown in the right panel. The zero velocity of all the spectra corresponds to CO(5-4) at $z=2.5$. The area filled with colors indicate the regions where positive emission is detected. 
\label{Fig:ALMA}}
\end{figure*}

In total, we detected three sources in CO(5-4) down to integrated line fluxes of $\sim$0.6 Jy km s$^{-1}$~(5$\sigma$) in the central 10$\arcsec$ region, two of which were also detected in CO(3-2). The spectroscopic redshifts of the three sources ($z_{spec} = 2.494, 2.503, 2.513$) are all consistent with being at the same structure at $z = 2.50$. The CO(5-4) spectra and intensity map are shown in Fig~\ref{Fig:ALMA}.

\subsection{VLT/K-band Multi-object Spectrograph (KMOS)}
Further near-infrared spectroscopic observation with the KMOS~\citep{Sharples:2004,Sharples:2013}) on the VLT was performed during P96 under ESO program 096.A-0891(A) (PI: Tao Wang) in 2015 December. KMOS is a multiplexed near-infrared integral-field spectroscopy (IFS) system with 24 deployable integral field units over a 7.2$\arcmin$ diameter field. The $K$-band filter was used to target the $H\alpha$ emission line for candidate members at $2 < z_{phot} < 3$ in and around the overdensity. The $K$-band filter covers the wavelength range $1.9-2.4$ $\mu$m, which corresponds to $H\alpha$ at $z \sim 1.9-2.6$. The spectral resolution in the $K$-band filter is around $R \sim 4200$.

\begin{figure*}[!tbh]
\centering
\includegraphics[scale=0.93]{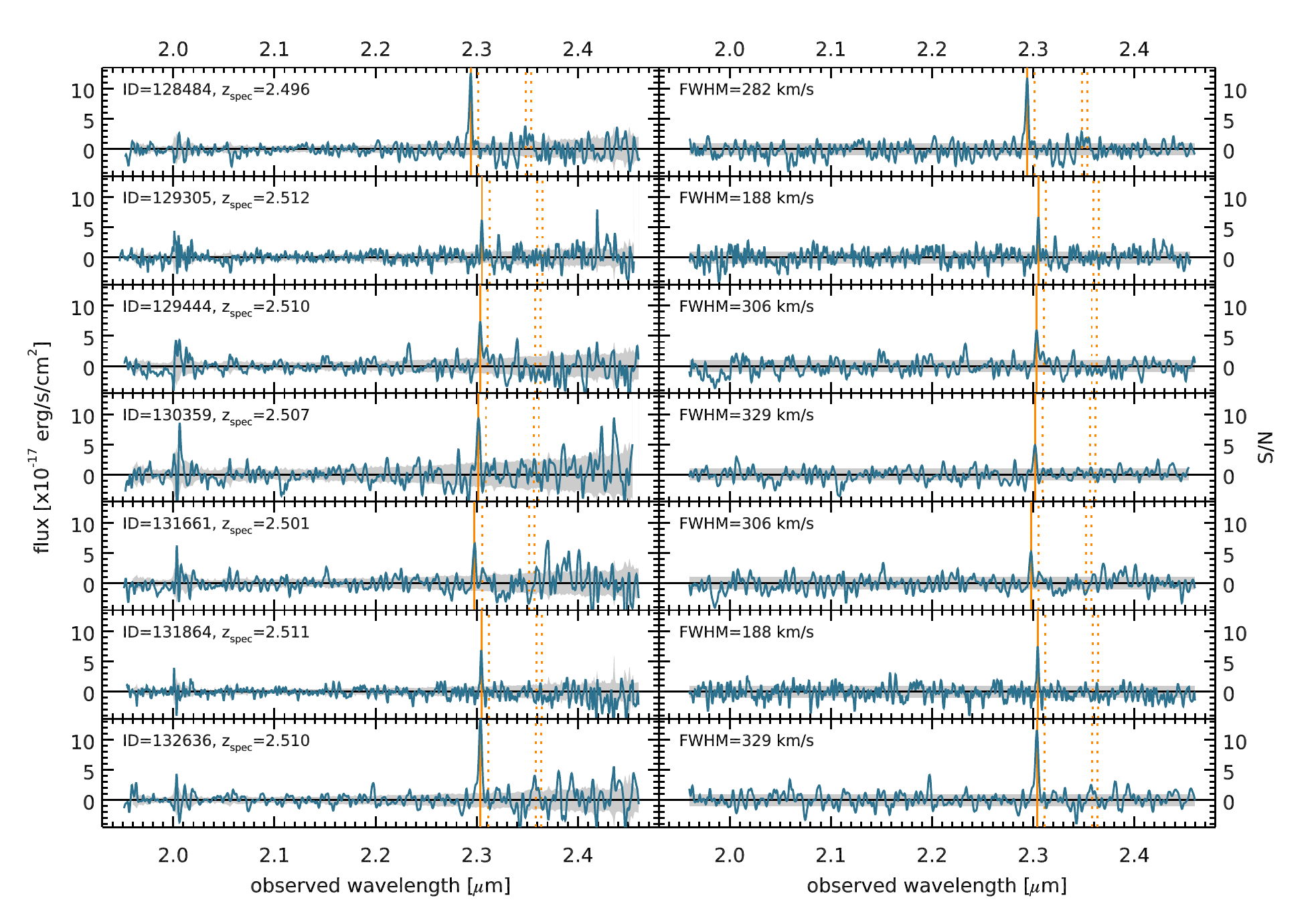}
\caption{Extracted and smoothed one-dimensional K-band KMOS spectra (left) and the corresponding S/N plot (right) for cluster members. The best-fit FWHM of each $H\alpha$ line is indicated in the right panels, which was used to smooth the spectra. All the listed sources are detected with a S/N $> 5$ except source 130359, which is detected at 4.9$\sigma$. Though in most cases we have only detected one line, given the range of their photometric redshifts (Table.~\ref{Tab:redshift})  we determined their redshifts assuming the line to be $H\alpha$. The position of the $H\alpha$ line is indicated by the solid line while dotted lines show the expected position of [NII] as well as the [SII] doublet.
\label{Fig:KMOS}}
\end{figure*}

The observations were prepared with the KMOS Arm Allocator (KARMA;~\citealt{Wegner:2008}) and each pointing was observed for 450 s using a standard object-sky-object dither pattern. The observations were taken in good conditions with a typical seeing of 0.6-0.8$^{\prime\prime}$.  The total on-source exposure time for each target is $\sim$ 1.5 hr. 
Data were reduced using the ESO pipeline (version 1.3.14) in combination with custom scripts developed by ourselves\protect\footnote{Detailed procedure and related codes are fully described here: https://github.com/cschreib/kmos-scripts}. 
We detected 11 galaxies with signal-to-noise ratio (S/N) above 4 in $H\alpha$ at $z > 2$ down to $f_{H\alpha} \sim 3 \times 10^{-17}$ erg s$^{-1}$ cm$^{-2}$. Seven of them are at $2.494 \leq z_{spec} \leq 2.512$, which are consistent with being cluster members. The spectra of these seven galaxies are shown in Fig.~\ref{Fig:KMOS}. 

\subsection{VLA}
Although our KMOS observation successfully detected a number of member galaxies, none of them are located in the core. This is likely caused by the fact that the DRGs in the core
are severely attenuated. To obtain spectroscopic redshifts for these massive and dusty sources and also constrain their molecular gas content, we performed CO(1-0) Karl G. Jansky Very Large Array (JVLA) observations of the cluster core in 2015 December under VLA program 15B-290 (PI: Tao Wang). Observations were carried out in the D array at the Ka band. The WIDAR correlation was configured with four spectral windows (SPWs) of 64 channels and 2 MHz per channel resolution. The effective frequency coverage is 32.2-33.59 GHz, corresponding to $z\sim 2.43-2.58$ for CO(1-0). 

\begin{figure}
\centering
\includegraphics[scale=0.65]{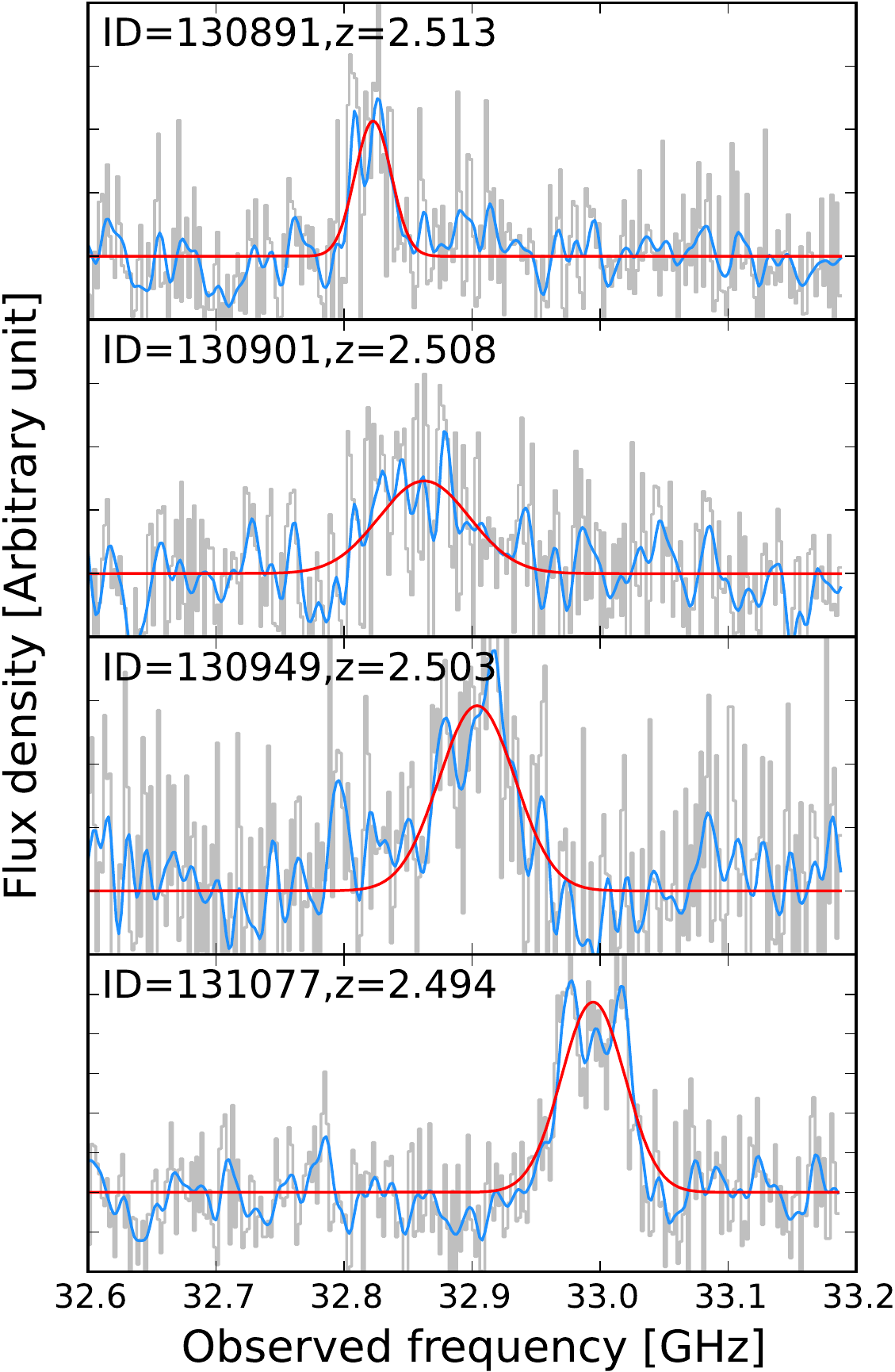}
\caption{Spectra of CO(1-0) emission for the four (out of five) ALMA detections that were detected in CO(1-0) with S/N $> 5$. The blue lines show the moving average of the spectra while the red lines show the best-fitting Gaussian profile. \label{Fig:VLA_spectra}}
\end{figure}

\begin{figure}[!tbh]
\centering
\includegraphics[scale=0.54]{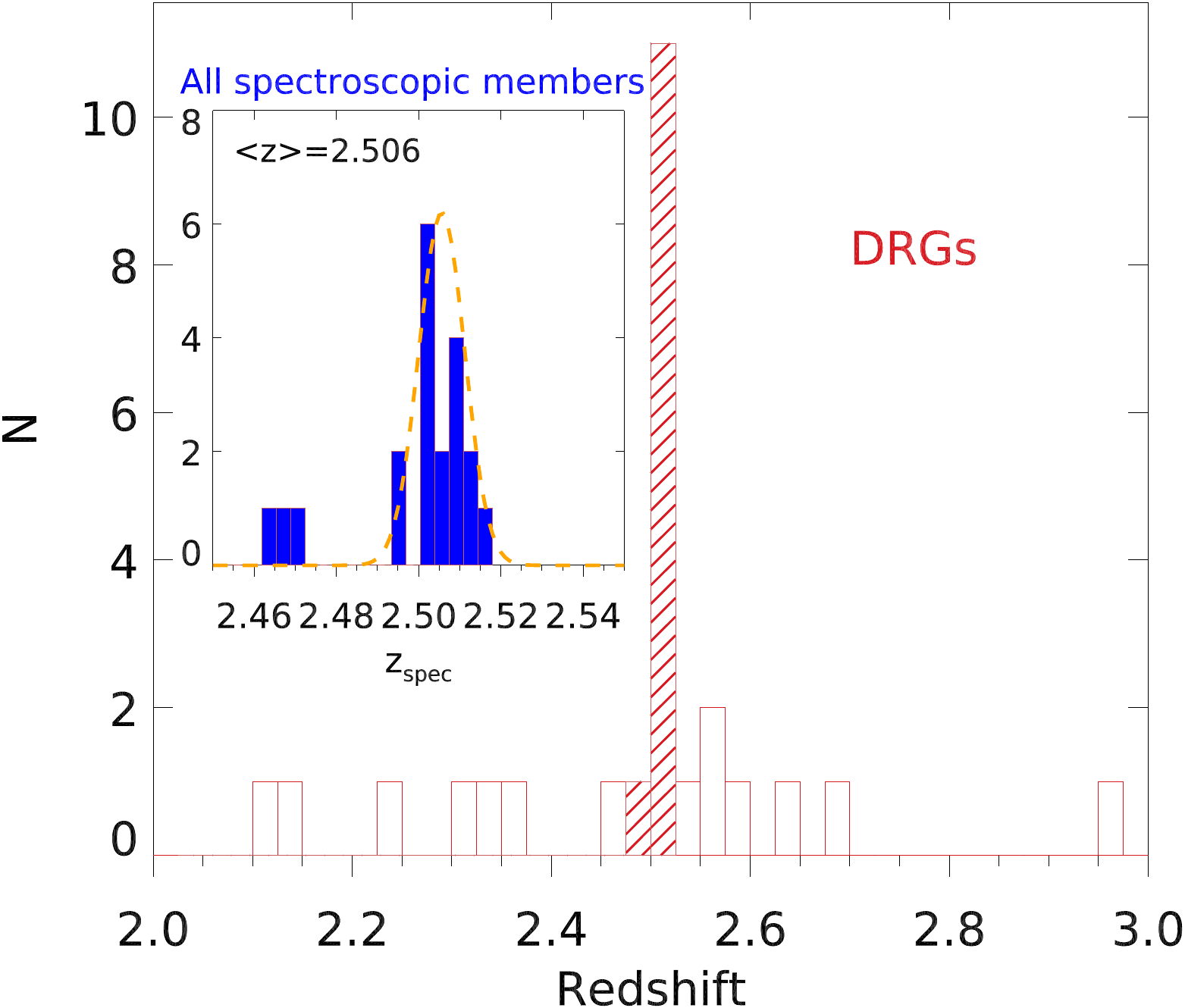}
\caption{Redshift distributions of the galaxies in and around the overdensity. The histogram in red shows the redshift distribution for all DRGs within 2$^\prime$ from the core 
of the structure. DRGs with spectroscopic redshift are indicated by the line-filled histogram. The inner panel shows the redshift distribution of all the galaxies in and around the overdensity (including non-DRGs) with spectroscopic redshift $2.45 < z_{spec} < 2.55$ from our spectroscopic surveys. 
\label{Fig:redshift}}
\end{figure}

The nearby quasar J1024-0052 was used for gain and pointing calibration and the source 3C147 was used as flux calibrator. The effective integration time is $\sim 10$ hr. 
The data were reduced using the Common Astronomy Software Application (CASA) package~\citep{McMullin:2007}, and were imaged using the CLEAN algorithm in CASA with a natural weighting scheme. More details on data reduction and molecular gas content of individual galaxies will be presented in a future work (T. Wang et al. 2016, in preparation). Eleven galaxies were detected in CO(1-0) down to integrated line fluxes 0.05 Jy km s$^{-1}$, including all the 3 CO(5-4) detections by IRAM/NOEMA and 1 $H\alpha$ detection by VLT/KMOS. Example CO(1-0) spectra of ALMA-detected galaxies in the cluster core are shown in Fig.~\ref{Fig:VLA_spectra}. Combining spectroscopic redshifts determined from IRAM-NOEMA and VLT/KMOS, we have in total spectroscopic redshifts for 21 galaxies extending up to $\sim$ 1 Mpc from the overdensity. Fig~\ref{Fig:redshift} shows the redshift distribution of these galaxies, which reveals a prominent spike at $z \sim 2.50$. The biweight mean of the redshifts of  these 21 galaxies yields $z_{mean} = 2.506$ with 17 galaxies falling in the range $z_{spec}=2.506\pm0.012$. The other five galaxies deviates from the mean by $> 3\sigma$. The redshift histogram distribution around the biweight mean is well described by a Gaussian profile. A maximum likelihood estimation of the dispersion with the 17 galaxies yielded $\sigma_{z} = 0.006$, and all the 17 galaxies fall in $z_{mean} \pm 3*\sigma_{z}$, hence are classified as cluster members. Spectroscopic redshifts of these 17 galaxies are listed in Table.~\ref{Tab:redshift}.

\section{$XMM-Newton$ and $Chandra$ imaging}
\label{Sec:Xray}
We combined the latest Chandra and $XMM-Newton$ surveys of the COSMOS field~\citep{Cappelluti:2009,Elvis:2009,Civano:2016}
to search for extended emission at a 16" spatial scale with wavelet-based detection techniques in the whole field. The depth of the Chandra and $XMM-Newton$
 survey reaches $\sim160$ ks and $\sim60$ ks per pointing, respectively. We co-added the $XMM$ and $Chandra$ background-subtracted count images 
and computed the total exposure. The procedure has been described in \cite{Finoguenov:2009} and shown to work even to much longer exposures such as those of CDFS in \cite{Finoguenov:2015}.
We used detailed background modeling  developed and verified in previous works~\citep{Cappelluti:2013,Erfanianfar:2013,
Finoguenov:2015}. A number of significant detections ($> 4\sigma$) associated with galaxy overdensities 
at high redshift were discovered (A. Finoguenov et al. 2016, in prep). The cluster reported in this paper exhibits the most prominent galaxy
overdensity which is also in the most advanced stage of follow-up observations.

\begin{figure*}[!tbh]
\centering
\includegraphics[scale=0.65]{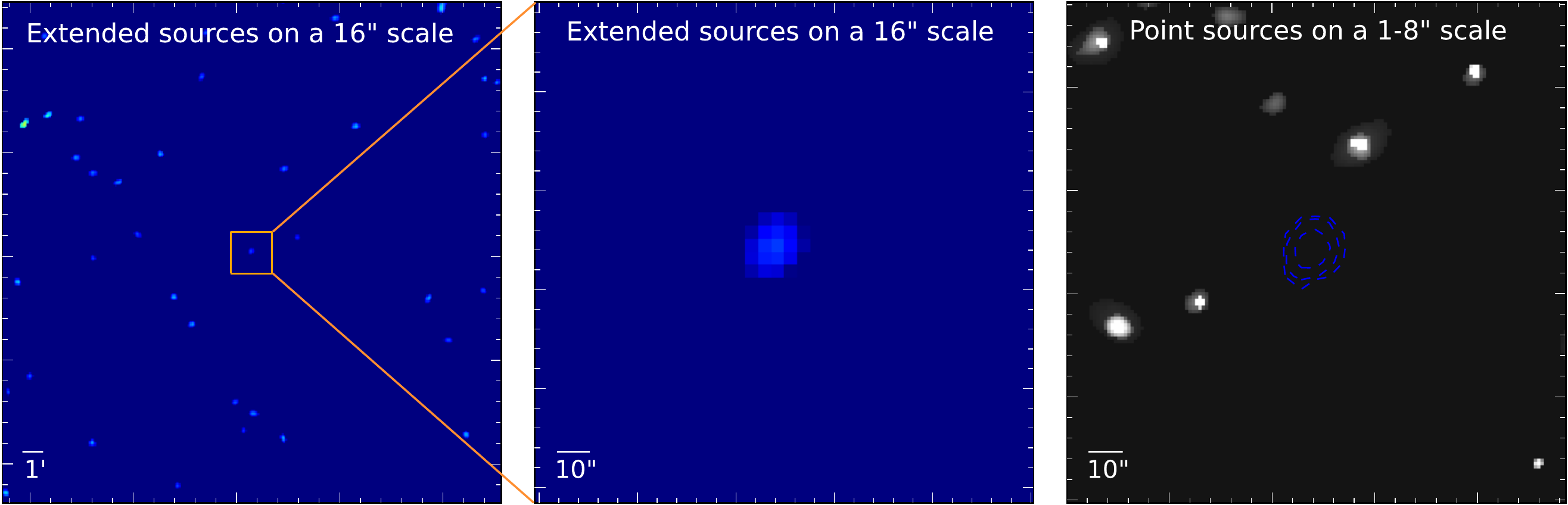}
\caption{X-ray detections ($> 4\sigma$ at $0.5-2$ kev) in the combined $Chandra$ and  XMM-Newton images on the 16" (left and middle) and 1-8" scales (right). The left panel shows a larger region ($0.4^\circ \times 0.4^\circ$)  while the middle and right panels show the central $2' \times 2'$ region around the cluster core. In the right panel, it is clear that the cluster does not have any significant detection at smaller scales. The blue contours indicate the extended emission of the cluster detected on the 16$^{\prime\prime}$ scale. 
\label{Fig:Xray_check}}
\end{figure*}

At the position of the cluster we did not detect any significant emission on scales smaller than 
16$\arcsec$ (Fig~\ref{Fig:Xray_check}), which eliminates the possibility of a point-source origin of the X-ray emission. 
With the selected detection threshold of 4$\sigma$, the possibility of detecting 
such a source by chance in the Chandra COSMOS survey can be rejected at 
99.5\% confidence. Roughly 90\% of the 4$\sigma$ detections are associated with a
galaxy overdensity (down to $K_{s} < 23.4$), with the one discussed in this paper being the most extreme case. 
The extended X-ray emission, together with the high number of matching spectroscopic redshifts, safeguards the 
detection from being a mere projection on the sky.

In order to derive the X-ray flux of the cluster, we used two apertures of radius 20$\arcsec$ and 32$\arcsec$. The 20$\arcsec$ aperture matches the extent of the detection while the 32$\arcsec$ aperture matches the expected size ($r_{500c}$) of the emission of the cluster, albeit with lower S/N. After a standard background subtraction,  we measured the net counts in the soft-band image ($0.5-2$ keV) for the small and large apertures, which are 57$\pm$15 and 78$\pm$23, respectively. To determine the total flux of the cluster, we started from the flux encompassing these two apertures and iteratively extrapolated them to $r_{500c}$. This iterative procedure is fully described in \cite{Finoguenov:2007}. In each step a correction factor is applied to the aperture flux assuming a $\beta$-model of the cluster brightness profile and an initial guess of $r_{500c}$, then a new $r_{500c}$ is estimated and a new correction factor is applied until it converges. The final estimate of $r_{500c}$ of the cluster is 24$\arcsec$ (or $\sim$185 kpc at $z=2.5$), and the total flux within $r_{500c}$ determined based on the two apertures are $5.4\pm1.4 \times 10^{-16}$ and $6.9\pm2.0 \times 10^{-16}$ erg~cm$^{-2}$~s$^{-1}$, respectively. The differences in the extrapolated estimates from the two apertures are attributed to the systematic errors and are small compared to the expected scatter in the $L_{X}-M_{halo}$ relation of $\sim$20\% in mass for our method~\citep{Allevato:2012}. Since the 32$\arcsec$ aperture is larger than the final estimate of $r_{500c}$, no flux extrapolation is needed. We hence take the flux measured from the 32$\arcsec$ aperture as our total flux estimate. This flux translates to an X-ray luminosity of $L_{0.1 - 2.4~\mathrm{keV}} = 8.8\pm2.6 \times 10^{43}$ erg s$^{-1}$ at $z=2.5$ with a $K$-correction from 0.5 to 2 keV to rest-frame 0.1-2.4 keV of 2.66. This X-ray luminosity measured in $r_{500c}$ is also used for calibrating the X-ray luminosity, $L_{X}$, to weak lensing mass at redshifts up to 1~\citep{Leauthaud:2010}.

\begin{table*}[!tbh]\centering
\begin{minipage}{\textwidth} 
\ra{1.3}
\caption{Measured physical properties of spectroscopically confirmed members\label{Tab:redshift}}
\begin{tabular}{@{}cccccccccc@{}}
\toprule
ID\footnote{IDs are from the $K_{s}$-selected catalog~\citep{Muzzin:2013a}} & RA  & DEC & $z_{\rm zpec}$  & $z_{\rm phot}$ & $J - K_{s}$ & log$M_{*}$   & Redshift determination  & Type \\ 
&  (J2000) &  (J2000) & & & & [$M_{\odot}$] &   &  & \\
\hline
128484 &  150.22348 & 2.30719 & 2.495 & 2.55   &  0.70   & 10.82    & $H\alpha$				& SF\\
129305 &  150.23940 & 2.31750 & 2.512 & 2.65   &  0.71   & 10.19    & $H\alpha$	                          & SF\\
129444 &  150.24875 & 2.31921 & 2.510 & 2.57   &  1.59   & 10.77    & $H\alpha$				& SF\\
131661 &  150.23584 & 2.34488 & 2.501 & 2.64   &  1.57   & 11.03    & $H\alpha$				& SF\\
131864 &  150.23454 & 2.34770 & 2.511 & 2.64   &  0.53   & 10.38    & $H\alpha$				& SF\\
132636 &  150.22505   & 2.35620 & 2.510 & 2.55 &  0.49   & 10.68    & $H\alpha$				& SF\\
130359 &  150.22899 & 2.32978 & 2.507 & 2.47   &  1.93   & 11.26	    & $H\alpha$, CO(1-0)    & SF\\
130842 &  150.23746   & 2.33612 & 2.515 & 3.04 &  1.77   & 11.12    & CO(1-0)				& SF\\
130891 &  150.23987   & 2.33645 & 2.513 & 2.68 &  2.09   & 11.06    & CO(1-0), CO(5-4) 	  	& SF\\
130901 &  150.23923   & 2.33637 & 2.508 & 2.20 &  1.74   & 11.58    & CO(1-0)	 			& SF\\
130933 &  150.23869   & 2.33683 & 2.504 & 2.28 &  2.23   & 11.29    & CO(1-0) 			    & SF (radio AGN)\\
130949 &  150.23701   & 2.33571 & 2.503 & 2.49 &  1.66   & 11.57    & CO(1-0), CO(3-2), CO(5-4) 	  	& SF\\
131077 &  150.23735   & 2.33814 & 2.494 & 2.82 &  1.39   & 11.16    & CO(1-0), CO(3-2), CO(5-4) 	 	& SF\\
131079 &  150.23695   & 2.33748 & 2.502 & 2.57 &  1.46   & 11.36    & CO(1-0)				& SF\\
132044 &  150.23650   & 2.34881 & 2.504 & 2.35 &  1.47   & 11.13	& CO(1-0)               & SF\\
132627 &  150.23421   & 2.35659 & 2.506 & 2.36 &  1.34   & 10.90	& CO(1-0)               & SF\\	
--\footnote{This source is clearly detected in the ultraVista images, but is close to a low-redshift galaxy hence was not included in the $K_{s}$-selected catalog.}	     &  150.23419  & 2.33647  & 2.504 & --     &  0.7  & 11.0    & CO(1-0)               & SF \\

\bottomrule
\end{tabular}
\end{minipage}
\end{table*}

While the non-detection at smaller scales suggests that the X-ray flux does not originate from a single galaxy, we investigate the possible contribution from combined star formation activities 
of the member galaxies. The aggregated infrared luminosity estimated from the combined infrared flux in the core, $10^{13.2} L_{\odot}$, translates to an X-ray luminosity $L_{0.5-2~\mathrm{kev}} = 1.26 \times 10^{43}$ erg s$^{-1}$ assuming the calibrated $L_{X}-SFR$ relation~\citep{Ranalli:2003}. Therefore we estimate that at most $\sim$14\% of the X-ray flux originates from star formation.

We also examine the possibility that the extended X-ray emission originated from inverse Compton (IC) scattering of the cosmic microwave background (CMB) photons by relativistic electrons in the central radio source. We argue that this is unlikely to be the case. The few structures/galaxies with extensive IC extended X-ray emission are mostly strong and extended radio galaxies (usually with strong jets,~\citealt{Fabian:2003,Overzier:2005,Johnson:2007}), with radio fluxes one to two orders of magnitudes higher than the central, point-like radio source of this structure. If we assume that the extended X-ray emission is fully produced through IC emission form the central radio source (ID 130933), we find that the resulting magnetic field is $\sim$0.5 $\mu$G~\citep{Harris:1979}, assuming a spectral slope of $-1$ for the radio emission. This magnetic field strength is significantly lower (radio emission is too weak compared to the X-ray emission) than the estimates for IC-origin structures in the literature, 30$\sim$180 $\mu$G~\citep{Overzier:2005}. Although there is still a possibility that the X-ray emission is due to an IC ghost/fossil jet of the central radio source (in this case, the source of the IC scattering is an older population of electrons, with a steeper spectral slope, from a previous outburst of the radio source), the non-detection of this structure at 324 MHz (3$\sigma$ upper limit $\sim$ 1.5 mJy,~\citealt{Smolvcic:2014}) suggests that the spectral index is less steep than $\sim-2$. Further deeper follow-up at low frequencies would be required to fully exclude this hypothesis, which is quite challenging with current facilities.

\begin{figure}[th]
\includegraphics[scale=0.46]{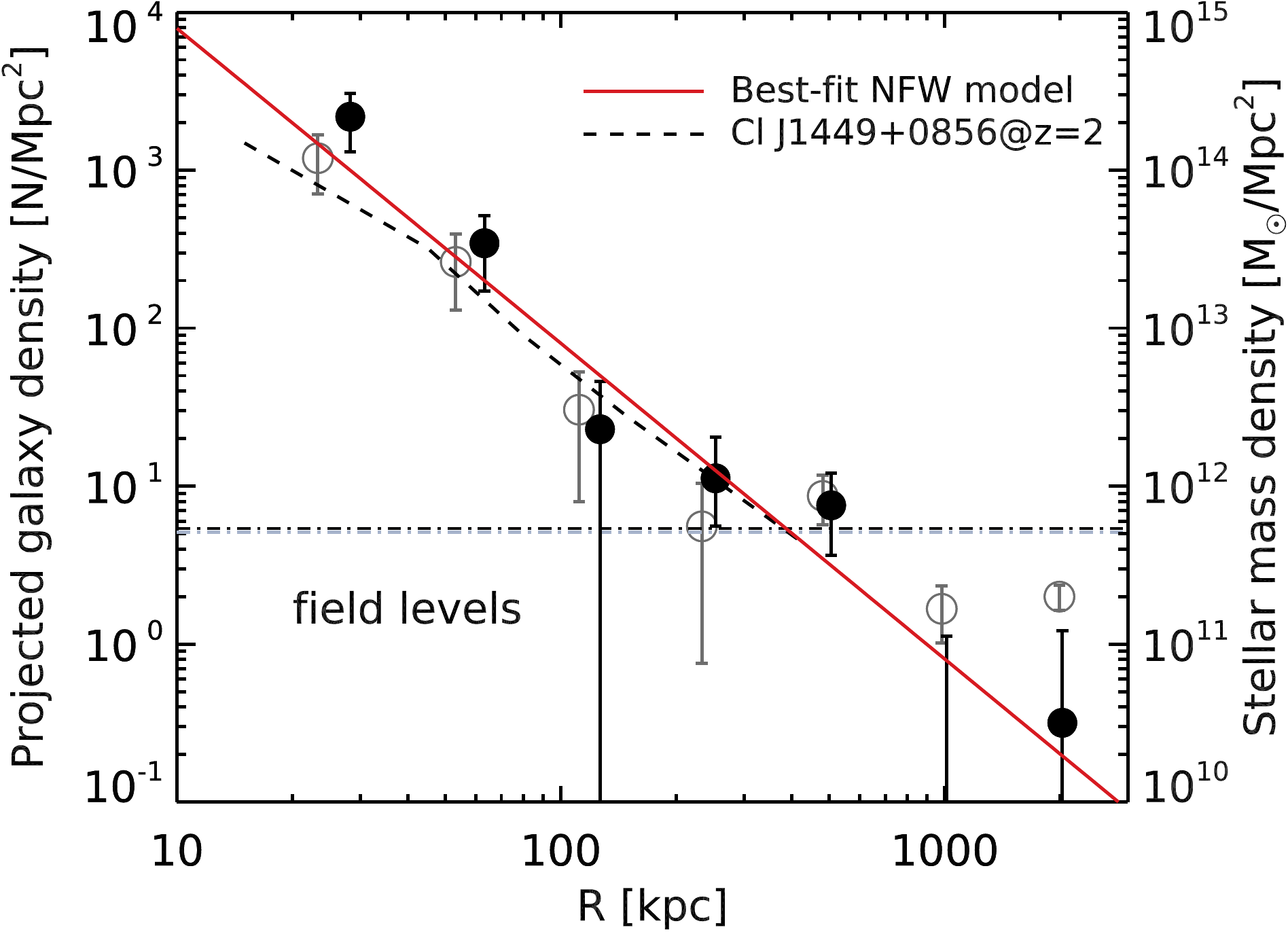}
\caption{Projected numbers (open circles) and mass densities (filled circles) of DRGs  extending to $\sim$ 2 Mpc from the overdensity. The average number and mass densities of field DRGs, as denoted by the gray and black horizontal lines, have been subtracted from the data points. The best-fitting projected NFW profile, which has a scale radius of $R_{s} = 0^{+8}_{-0}$ kpc, is shown by the red line. The stellar mass density profile of  the mature X-ray cluster Cl J1449+0856 at $z = 2$~\citep{Gobat:2011, Strazzullo:2013} is shown as the dashed line. 
\label{Fig:profile}}
\end{figure}

\section{CL J1001: A young galaxy cluster at $z=2.506$}
\label{Sec:cluster}
The high density of massive galaxies, extended X-ray emission and velocity dispersion of this overdensity suggest that it is embedded in a collapsed, cluster-sized halo, and hence a bona fide galaxy cluster. 
At a redshift of $z \sim 2.506$, it is the most distant, spectroscopically confirmed galaxy cluster, which pushes the formation time of galaxy clusters 
$\sim$0.7 Gyr back in cosmic time compared to previously discovered clusters~\citep{Gobat:2011}. In this section we further explore the structure and mass of CL J1001, and compare these properties to simulations.

\subsection{Structure and masses of CL J1001}
On large scales, this dense galaxy concentration is surrounded by a wider overdensity of DRGs at $z_{phot} \sim 2.50\pm0.35$ extending up to $\sim$1 Mpc (Fig.~\ref{Fig:profile}). The mass density profile of this structure, as calculated from these DRGs, resembles that of CL J1449+0856, a mature X-ray cluster with a total mass of  $M_{200c} \sim 6 \times 10^{13} M_{\odot}$ at $z \sim 2$~\citep{Strazzullo:2013}. The best-fitting projected NFW~\citep{Navarro:1997} profile of this structure is consistent with a single logarithmic slope of $-3$, suggesting that its host halo has a relatively high concentration, consistent with what has been observed in $z \sim 1$ clusters~\citep{vanderBurg:2014}. This provides further evidence of the existence of a virialized, cluster-sized halo of this structure. The absolute normalization of its mass density profile is most likely higher than Cl J1449+0856, since our $K_{s}$-selected catalog is only complete down to $M_{*} = 10^{11}M_{\odot}$ while the profile for Cl J1449+0856 was derived using a galaxy catalog complete down to $10^{10} M_{\odot}$.

We estimated the total halo mass of CL J1001 with three different methods, which are based on the X-ray luminosity, velocity dispersion, and the stellar mass content, respectively. 
Using the established $L_{X} - M_{halo}$ correlation in \cite{Leauthaud:2010}, the total X-ray luminosity of CL J1001, $L_{0.1-2.4 \rm{keV}} = 8.8\pm2.6 \times 10^{43}$ erg s$^{-1}$, corresponds to a 
total halo mass $M_{200c} \sim 10^{13.7\pm0.2} M_{\odot}$, which is comparable to that of the mature galaxy cluster CL J1449+0856 at $z \sim 2$ \citep{Gobat:2011}. While the redshift evolution in the $L_{X}-M_{halo}$ scaling relation used here is based on studies of clusters at $z < 1$, similar studies including more 
distant clusters yield consistent redshift evolution up to $z\sim1.5$~\citep{Reichert:2011}. The same scaling relation was also shown to be valid at $z \sim 2$ based on comparisons of the halo masses inferred from clustering analysis and stacked X-ray signals~\citep{Bethermin:2014}. Moreover, most of the detected 
deviations in the expected evolution in the $L_{X}-M_{halo}$  relation is driven by the evolution of the cool cores~\citep{Reichert:2011}. 
We assume 10\% of the emission to come from the cool core, which is typical for clusters studied in \cite{Leauthaud:2010}. 
Given the absence of the detection on smaller scales, a dominant contribution of 
the cool core to the total luminosity of this cluster can be ruled out. 

Galaxy cluster velocity dispersion provides another reliable tool for measuring cluster mass~\citep{Evrard:2008,Munari:2013,Saro:2013}. But we also note the large uncertainties in estimating mass for an individual cluster with velocity dispersion due to the influence of large-scale structure in and around clusters~\citep{White:2010}. The cluster redshift, $z = 2.506$, is determined by the biweight average of the 17 spectroscopic members. The galaxy proper velocities $v_{i}$ are then derived from their redshifts $z_{i}$ by $v_{i} = c(z_{i}-z)/(1+z)$ \citep{Danese:1980}. The line of sight velocity dispersion $\sigma_{v}$ is the square root of the biweight sample variance of proper velocities~\citep{Beers:1990,Ruel:2014}, which is estimated to be $\sigma_{v} = 530 \pm 120$ km s$^{-1}$. Using the relation between velocity dispersion and total mass suggested in \cite{Evrard:2008}, 
\begin{equation}
\sigdm(M,z) \ = \ \sigfifteen \ \biggl(\frac{h(z) \mtwoh}{10^{15}
  \Msol}\biggr)^\alpha  , 
\label{eq:DMVT}
\end{equation}
with $\sigfifteen$ the normalization at mass $10^{15} M_{\odot}$ and
$\alpha$ the logarithmic slope,  
we derived the total mass of CL J1001 to be $M_{200c} \sim 10^{13.7\pm0.2} M_{\odot}$ using the canonical value of $\sigfifteen \sim$ 1083 km s$^{-1}$ and $\alpha \sim 0.336$\footnote{We note that this relation is derived from DM-only simulations, however, simulations including baryonic physics yield fully consistent results~\citep{Munari:2013}.}. This estimate of the total mass is in good agreement with that derived from X-ray. We are aware that the sample used to estimate the velocity dispersion only includes SFGs. However, given that this cluster is dominated by SFGs (at at least $M_{*} > 10^{11} M_{\odot}$, where our sample is complete), we do not expect that including quiescent galaxies would change significantly the velocity dispersion estimation. Nevertheless, we are planning to spectroscopically confirm more member galaxies (including the quiescent ones) with follow-up observations, which will further improve the accuracy of the velocity dispersion estimation.

Studies of galaxy clusters at $z \sim 0-1.5$ show that the total stellar mass content is well correlated with the halo mass~\citep{vanderBurg:2014}, hence providing another tool to infer the cluster mass. 
To calculate the total stellar mass of the cluster we need to determine which galaxies are actual cluster members. Since not all the galaxies have spectroscopic redshifts, we have to rely on photometric redshifts determined by \cite{Muzzin:2013a}. 
Based on the 20 galaxies with spectroscopic redshifts at $2 < z < 3$, the normalized median absolute
deviation ($\sigma_\textsc{nmad}$\footnote{$\sigma_\textsc{nmad} = 1.48 \times \mathrm{median}\left( \left| \frac{\Delta z-\mathrm{median}(\Delta z)}{1+z_{spec}} \right| \right)$},~\citealt{Brammer:2008}) of $\Delta z = z_{phot}-z_{spec}$ is $\sigma _{NMAD}$ $\sim$ 0.033. Hence we define galaxies with redshifts with $ \left|z-2.506| \right/(1+2.506) <  3\sigma_\textsc{nmad}$ as candidate members. We added stellar masses for all the DRGs ($\sim50\%$ have spectroscopic redshifts) with this redshift range within $R_{200c}$ for a halo of $M_{200c} \sim 10^{13.7} M_{\odot}$ to determine the total stellar mass content. Field contamination is further estimated and subtracted based on the average surface number/mass density of DRGs in COSMOS at this redshift range.

The stellar mass of individual galaxies was derived from SED fitting with FAST~\citep{Kriek:2009a}. We fit the UV to 4.5 $\mu$m photometry with the \cite{Bruzual:2003} stellar population synthesis models, assuming solar metallicity and exponentially declining star formation histories with $e$-folding times $\tau \sim 0.1-10$ Gyr.  We allowed the galaxies to be attenuated with $A_{V}$ = 0 $-$ 6 with the Calzetti~\citep{Calzetti:2000} attenuation law. The mass estimate is in good agreement with those calculated by \cite{Muzzin:2013a} considering the small differences in the spectroscopic redshift used here and photometric redshifts used in their work. The median stellar mass of the 11  DRGs in the core is $\langle M_{*}\rangle \sim 10^{11.2} M_{\odot}$ with the two most massive ones reaching $M_{*} \sim 10^{11.6} M_{\odot}$. As a reference, for a DM halo of $M_{200c} \sim 10^{13} M_{\odot}$ (with a virial radius $R_{200c} \sim 186$ kpc) models expect only one galaxy as massive as $M_{*} \sim 10^{11.2} M_{\odot}$ at $z \sim 2.5$~\citep{Behroozi:2013a}, suggesting that this overdensity resides in a very massive halo.

We estimated the halo mass of this structure based on the stellar mass-to-total halo mass relation calibrated for $z \sim 1$ clusters~\citep{vanderBurg:2014}. If we add DRGs with $M_{*} > 10^{11} M_{\odot}$ (where our sample is 90\% complete), and do not apply any correction for mass incompleteness, we derive a combined stellar mass $M_{*} \sim 2.1 \times 10^{12} M_{\odot}$ after correction for field contamination, which is $< 10\%$. If instead we correct for the incompleteness and extrapolate down to $10^{9} M_{\odot}$ by assuming the same stellar mass function as that in the field as determined from the CANDELS fields~\citep{Grogin:2011,Koekemoer:2011,Schreiber:2015}, we derived a total stellar mass $M_{*} \sim 4.3 \times 10^{12} M_{\odot}$, suggesting $M_{200c} \sim 10^{14.6} M_{\odot}$. 
The true halo mass is most likely between the two estimates considering that massive galaxies tend to be more abundant in clusters than in the field (hence a smaller correction factor when extrapolated to lower stellar mass) as shown at $z \sim 1$~\citep{vanderBurg:2013}. Therefore, we estimate that the total mass of the halo is in the range of $M_{200c} \sim 10^{13.7-14.6} M_{\odot}$.  
Combined with the mass estimate based on X-ray and velocity dispersion, our best estimate of the halo mass is $M_{200c} = 10^{13.9 \pm 0.2} M_{\odot}$. 

 \begin{figure}[!tbh]
\includegraphics[scale=0.52]{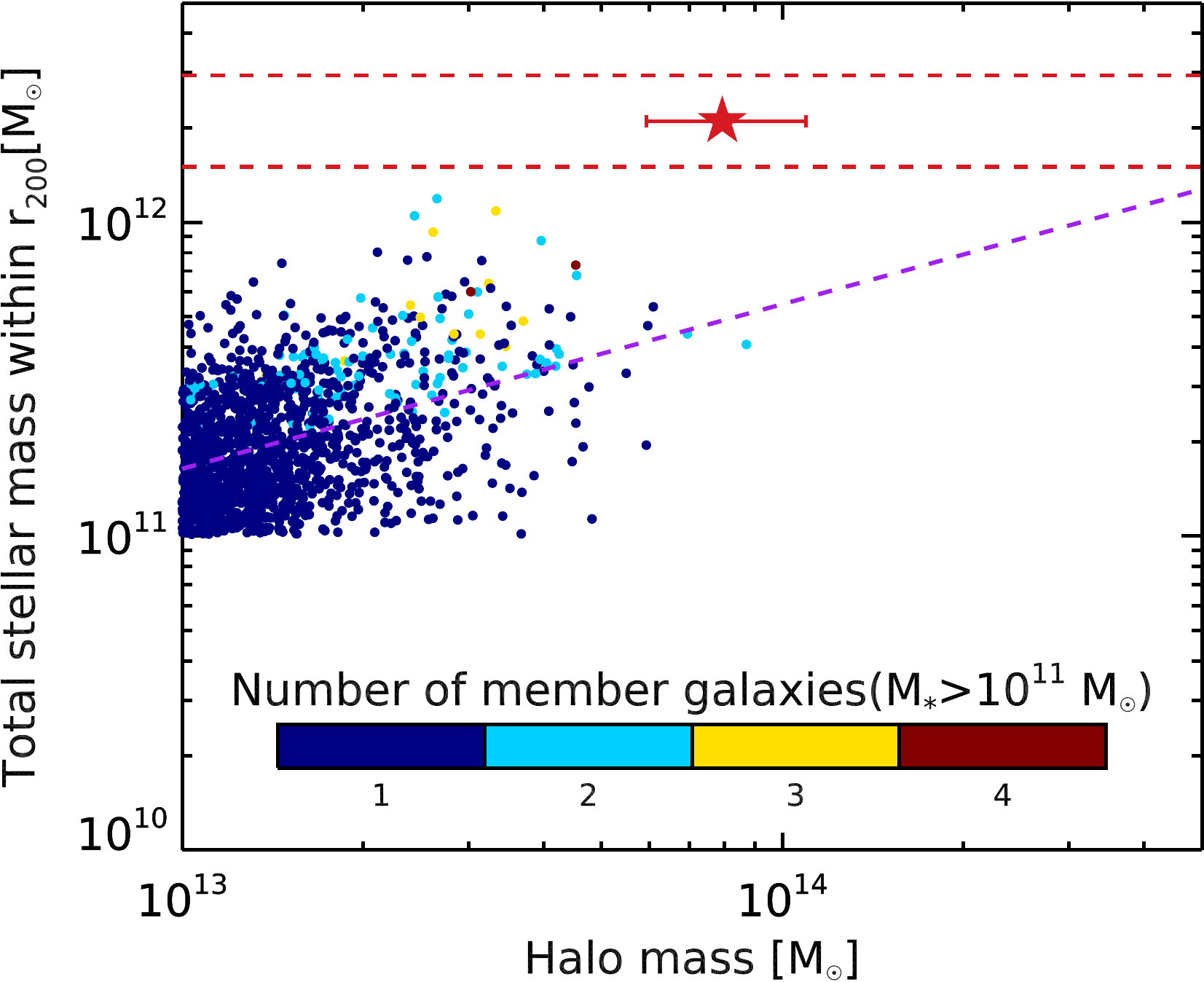}
\caption{Stellar mass content versus total halo mass for massive halos at $z=2.5\pm0.2$ from mock catalogs. The total stellar masses are calculated for galaxies with $K_{s} < 23.4$ mag and $M_{*} > 10^{11} M_{\odot}$ within $R_{200c}$ of the halo center, as defined as the most massive galaxy in each halo. The data point for each halo is color-coded by the number of member galaxies with $M_{*} > 10^{11} M_{\odot}$. The purple dashed line indicates the best linear fit of the stellar mass-to-halo mass relation based on the data points. The stellar mass estimate for CL J1001 and its uncertainties are denoted by the red dashed lines. The red star indicates our estimate of the halo mass and total stellar mass (also only accounting for galaxies with $M_{*} > 10^{11} M_{\odot}$) for CL J1001. 
\label{Fig:simulation}}
\end{figure}

\subsection{Comparison with simulations}

We conclude that we have found a massive galaxy concentration embedded in a virialized, cluster-sized halo at $z=2.506$. 
Halos of similarly high masses at these redshifts are predicted to be very rare in the $\Lambda$-CDM framework. The cumulative number of DM halos with $z > 2.5$ and $M_{200c} > 1(0.5) \times 10^{14}M_\odot$ expected in the COSMOS field is  0.01(0.3) with Planck cosmology (twice lower using WMAP 7 cosmology)~\citep{Murray:2013,PlanckXIII:2015}. More accurate halo mass estimates and more similarly massive structures at high redshifts are hence required to put stringent constraints on our cosmological model.

To understand properties of this cluster in a cosmological context, we have searched similar structures in mock catalogs for light cones ~\citep{Henriques:2012,Henriques:2015} constructed for the semi-analytic galaxy formation simulation of \cite{GuoQ:2011}, which were built on merger trees from large DM simulations, the Millennium Simulation~\citep{Springel:2005} and Millennium-II~\citep{Boylan-Kolchin:2009} Simulation. These mock catalogs include in total a $\sim$47 times larger area than the COSMOS field. We have extracted all the friends-of-friends (FoF) halos at $z = 2.5\pm0.2$ exceeding $M_{200c} = 10^{13} M_{\odot}$. We then derived the total stellar mass of galaxies with $M_{*} > 10^{11} M_{\odot}$ in each halo, and present this total stellar mass versus halo mass in Fig.~\ref{Fig:simulation}. The stellar mass content of CL J1001 appears significantly higher, by a factor of $\sim 4$, than those of similarly massive halos in simulations. This may indicate that the stellar mass build-up in massive clusters at $z > 2$, at least for the most massive galaxies, is more rapid than what was predicted in simulations and semi-analytical models. Alternatively, it may suggest that structures like CL J1001 are extremely rare and are not present even in today's largest cosmological simulations.

Based on the evolution of the mass of the most massive progenitor halo in simulations~\citep{Springel:2005,Chiang:2013}, a massive halo with $M_{200c} \gtrsim 10^{13.9} M_{\odot}$ at $z = 2.5$ will evolve into a  $\sim2 \times 10^{15} M_{\odot}$ halo at $z=0$, i.e., a ``Coma"-type cluster ($M_{200c} > 10^{15} M_{\odot}$). As shown in these simulations, the main progenitor (the most massive halo) of a  ``Coma"-type  cluster at $z > 2$ is embedded in a large-scale filamentary structure spreading over several tens of Mpc. Hence if CL J1001 will evolve into a ``Coma"-type cluster at $z \sim 0$, then the presence of a large-scale galaxy structure around its position is expected. Indeed, a large-scale filamentary structure at $z \sim 2.47$ whose geometric center coincides with CL J1001 was recently spectroscopically confirmed in the same field~\citep{Casey:2015}. The same region has also been identified as a protocluster candidate at similar redshifts based on photometric redshifts~\citep{Chiang:2014}. The redshift difference (2.47 versus 2.50) corresponds to a comoving line of sight distance of $\sim$ 35 Mpc, which is consistent with the extension of the progenitor of a massive cluster at these redshifts. 
The presence of this $\sim$30 Mpc scale overdensity surrounding CL J1001 provides further evidence that it will eventually form a massive, ``Coma"-type galaxy cluster in the present day Universe. 

 \begin{figure}
\includegraphics[scale=0.55]{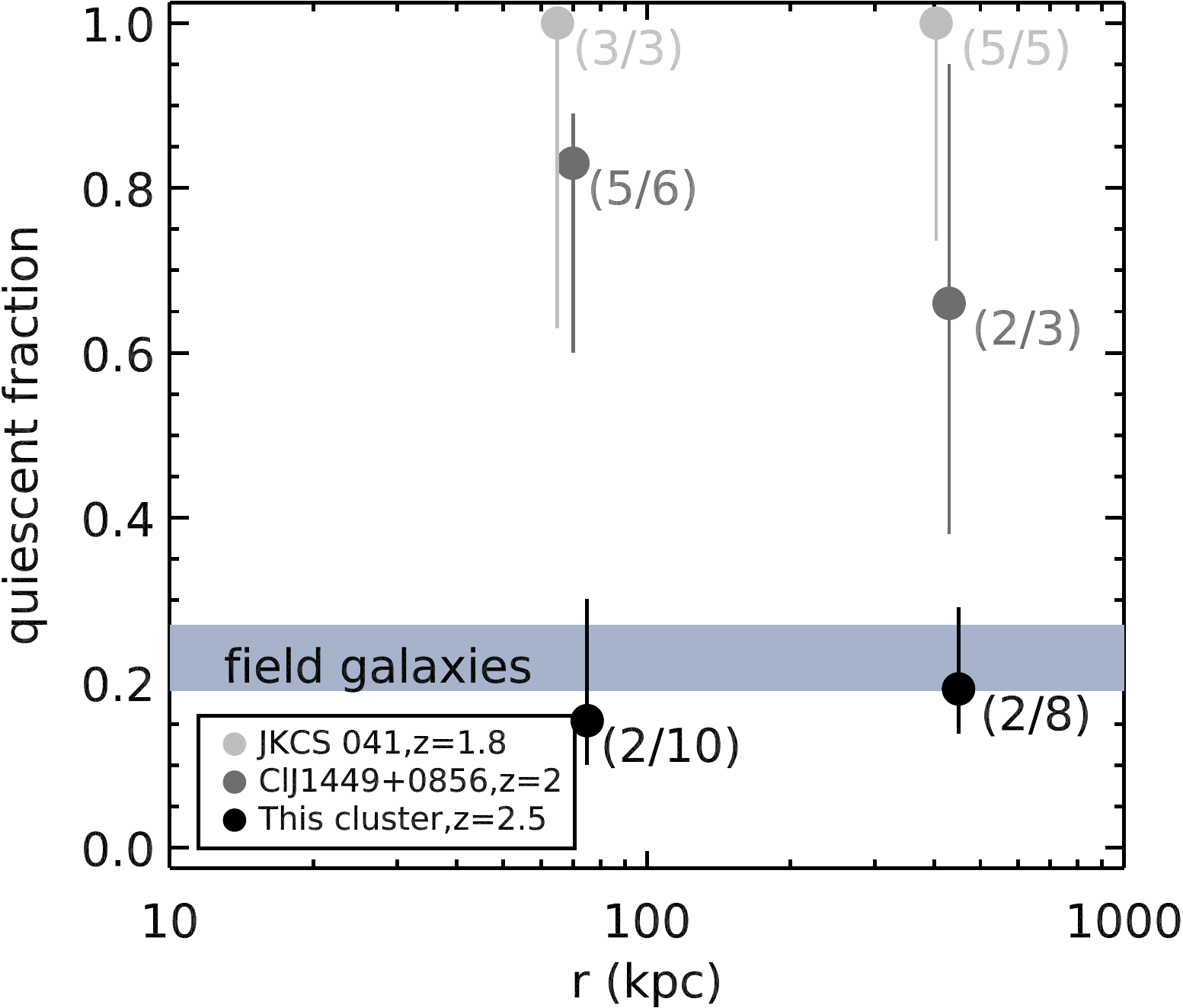}
\caption{Fraction of galaxies classified as quiescent as a function of clustercentric radius. The quiescent fraction at $M_{*} > 10^{11} M_{\odot}$ in this structure as well as two mature clusters at $z \sim 2$ (CI J1449+0856 and JKCS 041) are shown. The quiescent fraction is calculated for two radial bins of clustercentric radius ($r < 150$ kpc and 150~$< r < 700$ kpc), with the number of galaxies used to calculate this fraction indicated at each data point. Error bars are 1$\sigma$ confidence level for binomial population proportions. The shaded region indicates the quiescent fraction at the same stellar mass at $z=2.5\pm0.3$ for field galaxies, which was derived from the CANDELS and 3D-HST~\citep{Brammer:2012} survey.  
\label{Fig:passive_frac}}
\end{figure}

\section{Properties of member galaxies} 
\label{Sec:galaxies}

\subsection{Star formation and  Supermassive Black Hole Accretion in the Cluster Core}

\begin{table*}[!tbh]\centering
\begin{minipage}{\textwidth} 
\ra{1.3}
\caption{Infrared and radio properties of member galaxies detected with ALMA\label{tab:IR}}
\begin{tabular}{@{}cccccccccccc@{}}
\toprule
ID\footnote{IDs are from the $K_{s}$-selected catalog~\citep{Muzzin:2013a}}  & $S_{24 \mu m}$ & $S_{100 \mu m}$ & $S_{160 \mu m}$ & $S_{250 \mu m}$ & $S_{350 \mu m}$ & $S_{500 \mu m}$ & $S_{870 \mu m}$\footnote{flux densities for individual galaxies at 870 $\mu$m are from~\cite{Bussmann:2015} while the combined flux (at a slightly different wavelength, 850~$\mu$m) in the core is from~\cite{Casey:2013}}  & $S_{1.1 mm}$ & $S_{1.8 mm}$ & $S_{1.4 GHz}$ & log$L_{IR}$\\
  & [mJy] & [mJy] & [mJy] & [mJy]  & [mJy] & [mJy] & [mJy]  & [mJy] & [mJy] & [$\mu$Jy] & [$L_{\odot}$] \\
\hline
130891  & 0.15$\pm$0.01 & $<4.5$ 	  & $<9.8$     & --       & --       & --       & 3.77$\pm$0.32  & --        & 0.39$\pm$0.07 & 40$\pm$13 & 12.6$\pm$0.15\\ 
130901  & 0.04$\pm$0.01 & $<4.5$ 	  & $<9.8$     & --       & --       & --       & 1.66$\pm$0.21  & --        & $<0.2$        & $<40$     & 12.0$\pm$0.15\\ 
130933  & 0.06$\pm$0.01 & $<4.5$ 	  & $<9.8$     & --       & --       & --       & 2.23$\pm$0.41  & --        & 0.24$\pm$0.07 & 81$\pm$15 & 12.2$\pm$0.15\\ 
130949  & 0.17$\pm$0.02 & $<4.5$ 	  & $<9.8$     & --       & --       & --       & 1.69$\pm$0.25  & --        & 0.09$\pm$0.07 & $<40$     & 12.5$\pm$0.15\\ 
131077  & 0.28$\pm$0.02 & 6.3$\pm$1.7 & 20.1$\pm$2 & --       & --       & --       & 5.26$\pm$0.26  & --        & 0.63$\pm$0.07 & 78$\pm$13 & 12.8$\pm$0.10\\ 
cluster core    & --            & --          & --         & 61$\pm$6 & 77$\pm$6 & 66$\pm$6 & 14.8$\pm$2     & 8.9$\pm$2 & --            & --        & 13.2$\pm$0.1\\ 

\bottomrule
\end{tabular}
\end{minipage}
\end{table*}

While the halo mass and peak galaxy density of CL J1001 already resemble those of low-redshift mature clusters, member galaxies in the core of CL J1001 show unusual star formation properties compared to previously discovered clusters, as indicated by the exceptional far-infrared (FIR) and millimeter emission in the core. Using the total infrared flux densities measured in the core, we conducted an infrared SED fitting to derive its infrared luminosity in order to estimate the combined SFRs of the massive galaxies in the core. To keep the SED fitting simple, we fit the 250 $\mu$m to 1.1mm data points with an FIR SED consisting of a coupled single-dust-temperature blackbody and mid-infrared power law described in \cite{Casey:2012a}. We have also fitted the SED with the \cite{Chary:2001} infrared SED templates. The two methods yielded similar results,  $L_{IR} = 10^{13.2\pm0.1} L_{\odot}$. 
This infrared luminosity translates to an SFR of $\sim 3400 M_{\odot}$ yr$^{-1}$ based on the calibration in \cite{Kennicutt:1998}. Such high SFRs were seen in some of the protoclusters at high redshifts~\citep{Daddi:2009a},however, these are quite unusual for a structure with such a high concentration of massive galaxies, whose peak massive galaxy density is already comparable to that of mature clusters at lower redshifts.

\begin{figure}[!tbh]
\centering
\includegraphics[scale=0.46]{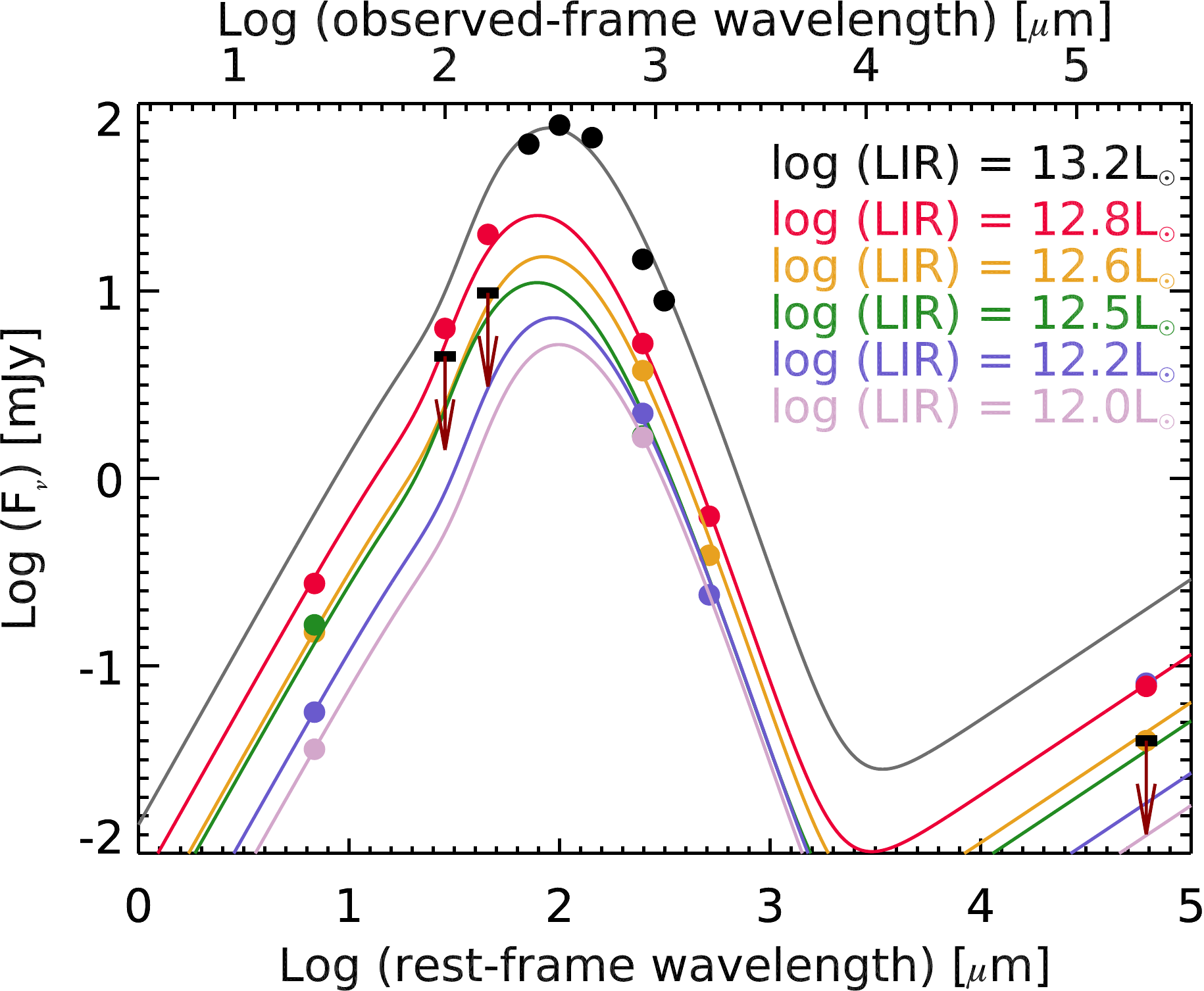}
\caption{Infrared SEDs for the five candidate members with ALMA 870 $\mu$m detections. The colored solid lines show the SED fitting results for the 5 ALMA detections while the black line shows the SED fitting result for the combined flux densities in the core. The 3$\sigma$ upper limits for non-detections at 100 $\mu$m, 160 $\mu$m, and 1.4 GHz are shown.
\label{Fig:IR-SED}}
\end{figure}

We also derived the dust mass in the cluster core from the best-fit graybody templates of the combined SED. During the fit, the dust emissivity index is fixed at $\beta = 1.5$ to reduce the number of free parameters. We then converted dust masses into gas masses assuming a metallicity determined from the mass-metallicity relation at $z = 2.5$ ~\citep{Erb:2006}. For the conversion we assumed a metallicity corresponding to a galaxy with $M_{*} \sim 10^{11.2} M_{\odot}$, the median mass of the 11 DRGs in the core. The total dust mass is then calculated to be  $M_{dust} = 10^{9.3\pm0.1} M_{\odot}$, translated to a molecular gas mass of $M_{gas} = 5\pm1 \times 10^{11} M_{\odot}$. This large amount of molecular gas and the high current SFR suggest that the cluster core is still actively assembling its stellar mass and will increase its mass substantially in a short timescale despite the fact that its current stellar mass density (or the number of massive galaxies) is already comparable to that of mature clusters.

Only 2 out of the 11 DRGs in the 80kpc core are classified as quiescent galaxies based on the rest-frame $UVJ$ diagram~\citep{Williams:2009}. Figure~\ref{Fig:passive_frac} shows the quiescent fraction at $M_{*} > 10^{11} M_{\odot}$ as a function of clustercentric radius for massive galaxies in CL J1001, as well as for two X-ray clusters, ClJ1449+0856 and JKCS 041 at $z \sim 2$. Unlike ClJ1449+0856 and JKCS 041, which show a quiescent fraction of 70\%-100\%, the quiescent fraction in CL J1001 is significantly lower, $\sim 20\%$, indistinguishable from that in the field at the same redshifts. 

 \begin{figure}
\centering
\includegraphics[scale=0.5]{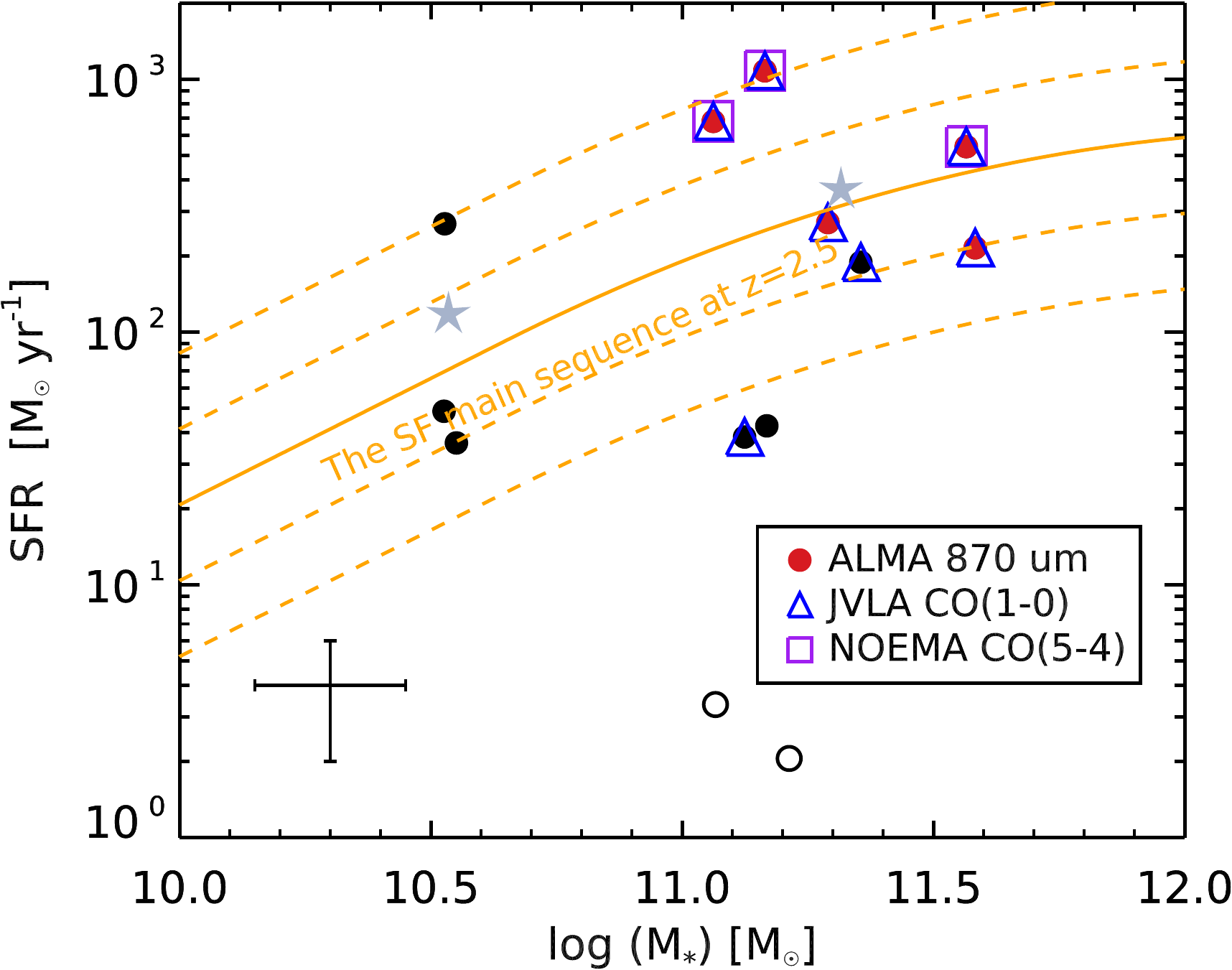}
\caption{SFR-stellar mass relation for the 13 member galaxies in the cluster core (within 10$^{\prime\prime}$). Filled circles denote galaxies that are classified as star-forming, while open circles are galaxies classified as quiescent based on the rest-frame $U - V$ versus $V - J$ diagram. Galaxies that are detected at 870 $\mu$m with ALMA, CO(1-0) with JVLA, and CO(5-4) with IRAM-NOEMA are respectively denoted by red filled circles, blue open triangles,  and purple open squares. SFRs for ALMA 870 $\mu$m detected sources are derived from their infrared luminosity while the others are from UV to NIR SED fitting. The main sequence of field star-forming galaxies at $z \sim 2.5$ and associated 0.3 dex and 0.6 dex scatter~\citep{Schreiber:2015} are shown with orange lines. The filled stars indicate the mean value for the star-forming members in two mass bins separated at $M_{*} = 10^{11} M_{\odot}$.
\label{Fig:MS_check}}
\end{figure}

The rich available data set in this structure allows us to probe further star formation properties of individual galaxy members. In particular, for the five ALMA detections, we performed a prior-based PSF-fitting using FASTPHOT~\citep{Bethermin:2010} at shorter wavelengths, i.e., 24, 100, and 160~$\mu$m to obtain the full far-infrared SEDs. 
We then estimated $L_{IR}$ by fitting their infrared SEDs across 24$\mu$m, 100$\mu$m, 160$\mu$m, 870 $\mu$m, and 1.8 mm bandpasses using the same method as we did for the combined SED. Fig.~\ref{Fig:IR-SED} shows the best-fit infrared SEDs for the five galaxies detected with ALMA, as well as the combined infrared and millimeter emission in the cluster core. The derived $L_{IR}$ are listed in Table.~\ref{tab:IR}. All the five ALMA detections have $L_{IR} > 10^{12} L_{\odot}$. Based on the \cite{Kennicutt:1998} relation, the $L_{IR}$ of these five galaxies adds up to an SFR $\sim 2700 M_{\odot}$ yr$^{-1}$, which is consistent with our SFR estimation using the combined infrared SED (3400 $M_{\odot}$ yr$^{-1}$) considering that some of the star-forming members are not detected with ALMA. For non-ALMA detected sources in the cluster core, we estimated their SFR based on SED fitting results with FAST. 

Figure~\ref{Fig:MS_check} presents the stellar mass-SFR relation for galaxies in CL J1001 and their comparison with field galaxies. The two brightest ALMA detections, or 25\%(2/8)\footnote{If we exclude the two $UVJ$-SFGs that fall 0.6 dex below the star formation main sequence, the starburst fraction would be 33\%.}  in terms of fraction, have SFR $\sim 4$ times higher than the main-sequence SFGs at the same mass, indicating an elevated starburst activity in this structure (this starburst fraction is $\sim 3-5\%$  in field galaxies;~\citealt{Elbaz:2011,Rodighiero:2011,Schreiber:2015}). Averaging over all the star-forming members (as shown by the filled stars in Figure~\ref{Fig:MS_check}), the mean SFR versus stellar mass relations for these cluster SFGs fall on the same relation within uncertainties as galaxies in the field.

Moreover, 3 out of the 11 DRGs are detected at 1.4 GHz with $F_{1.4 GHz} \sim 70-80~\mu$Jy though none are detected in X-rays. Two of them are classified as radio AGNs based on their larger radio-to-IR luminosity ratio than that for normal SFGs. This (radio) AGN fraction ($\sim$18\%) is is much higher than that in the field ($\lesssim 3.8\%$, assuming that all the DRGs with 1.4 GHz detections down to $F_{1.4 GHz} \gtrsim 50 \mu$Jy are radio AGNs), suggesting an enhanced radio AGN activities in this dense structure.

\subsection{Structural properties of member galaxies}
We studied structural properties of cluster member galaxies using HST/WFC3 $J_{110}$ image (711.74 s integration time) from the \textit{HST} archive (PI: M. Negrello). Data were reduced using the IRAF MultiDrizzle package (see ~\citealt{Negrello:2014} for further details). With the algorithm GALFIT~\citep{PengC:2010} we fitted the galaxy light distribution with a single S\'ersic law~\citep{Sersic:1968}. Uncertainties associated with $r_{\mathrm{e}}$ measurements were derived through Monte Carlo simulations by fitting simulated galaxies that were injected into the real image. 

\begin{figure}[!tbh]
\includegraphics[scale=0.5]{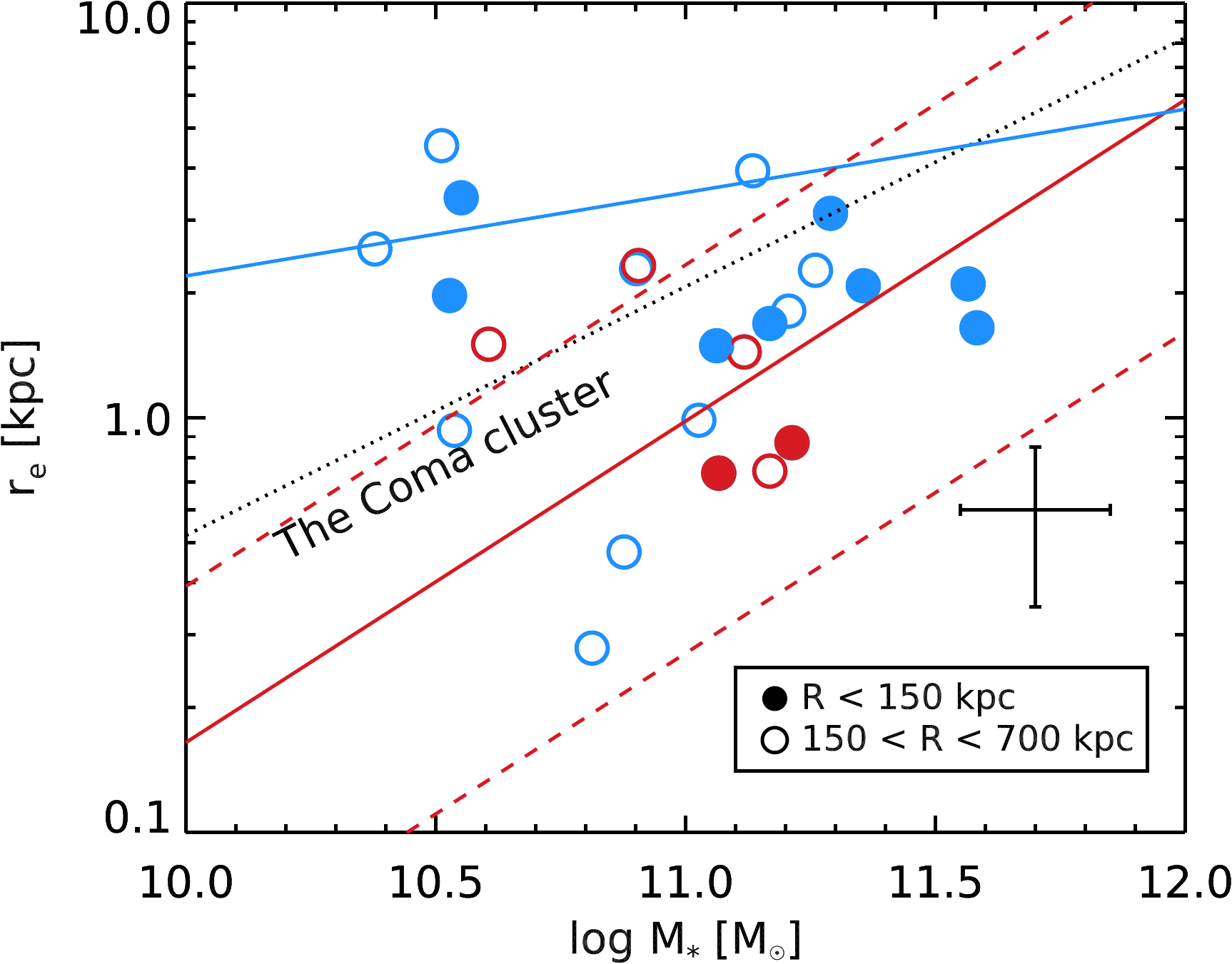}
\caption{Mass-size relation for quiescent and star-forming candidate members of the structure. Galaxies within $R < 150$ kpc from the core are denoted by filled circles while those within 150 $< R <$ 700 kpc are denoted by open circles. $UVJ$-quiescent galaxies are shown in red, while $UVJ$-star-forming galaxies are shown in blue. The mass-size relations for quiescent and star-forming field galaxies at $z \sim 2.5$~\citep{vanderWel:2014} are shown with red and blue lines, respectively. The 1$\sigma$ scatter of the mass-size relation for field quiescent galaxies is shown with red dashed lines. The large error bar in the bottom right indicates typical uncertainties of mass and size measurements of our sample galaxies. The mass-size relation for the nearby Coma cluster is indicated with the dotted line\citep{Andreon:1996}.  
\label{Fig:mass_size}}
\end{figure}

The $J_{110}$ band corresponds to rest-frame $\sim 3300$~\AA~at $z \sim 2.5$. To make a proper comparison with other studies, we corrected this 
size to rest-frame 5000~\AA~following an empirically calibrated morphological $k$-correction relation for quiescent and SFGs in the CANDELS fields~\citep{vanderWel:2014}. Both corrections are relatively small and our conclusion remains unchanged with and without applying this morphological $k$-correction.

We show the mass-size relation of candidate cluster members in Fig.~\ref{Fig:mass_size}. At $M_{*} > 10^{11} M_{\odot}$ both quiescent and star-forming members fall on the mass-size relation for field quiescent galaxies, which are more compact than their local counterparts. The fact that the quiescent members in CL J1001 are as compact as those in the field differs from what has been found in a number of mature clusters at $z \sim 1-2$~\citep{Papovich:2012,Bassett:2013,Strazzullo:2013,Delaye:2014,Newman:2014}, in which the quiescent galaxies are less compact than their field counterparts (so-called accelerated evolution of the mass-size relation in clusters). This is consistent with the fact that CL J1001 was caught in an earlier phase of cluster formation (right after the collapse of the cluster-sized halo), when the cluster environment had not yet affected the structural evolution.  On the other hand, in contrast to the mass-size relation for SFGs in the field, most of the massive star-forming members in CL J1001 are significantly smaller and  fall on the same mass-size relation as quiescent galaxies. This indicates that these cluster SFGs are promising progenitors of quiescent galaxies and may soon be quenched.

\section{Discussion: Implications of CL J1001 on the formation of massive clusters and their member galaxies}
\label{Sec:discussion}
With the presence of both a massive, collapsed halo and a predominant population of massive SFGs, CL J1001 provides a rare chance to  
study the rapid build-up of a dense cluster core. The discovery of structures in such a phase itself helps to bridge the gap between previously discovered photoclusters and clusters at high redshift. Its properties provide new insights into when and how massive cluster ellipticals formed at high redshift.    

\begin{figure}[!tbh]
\includegraphics[scale=0.365]{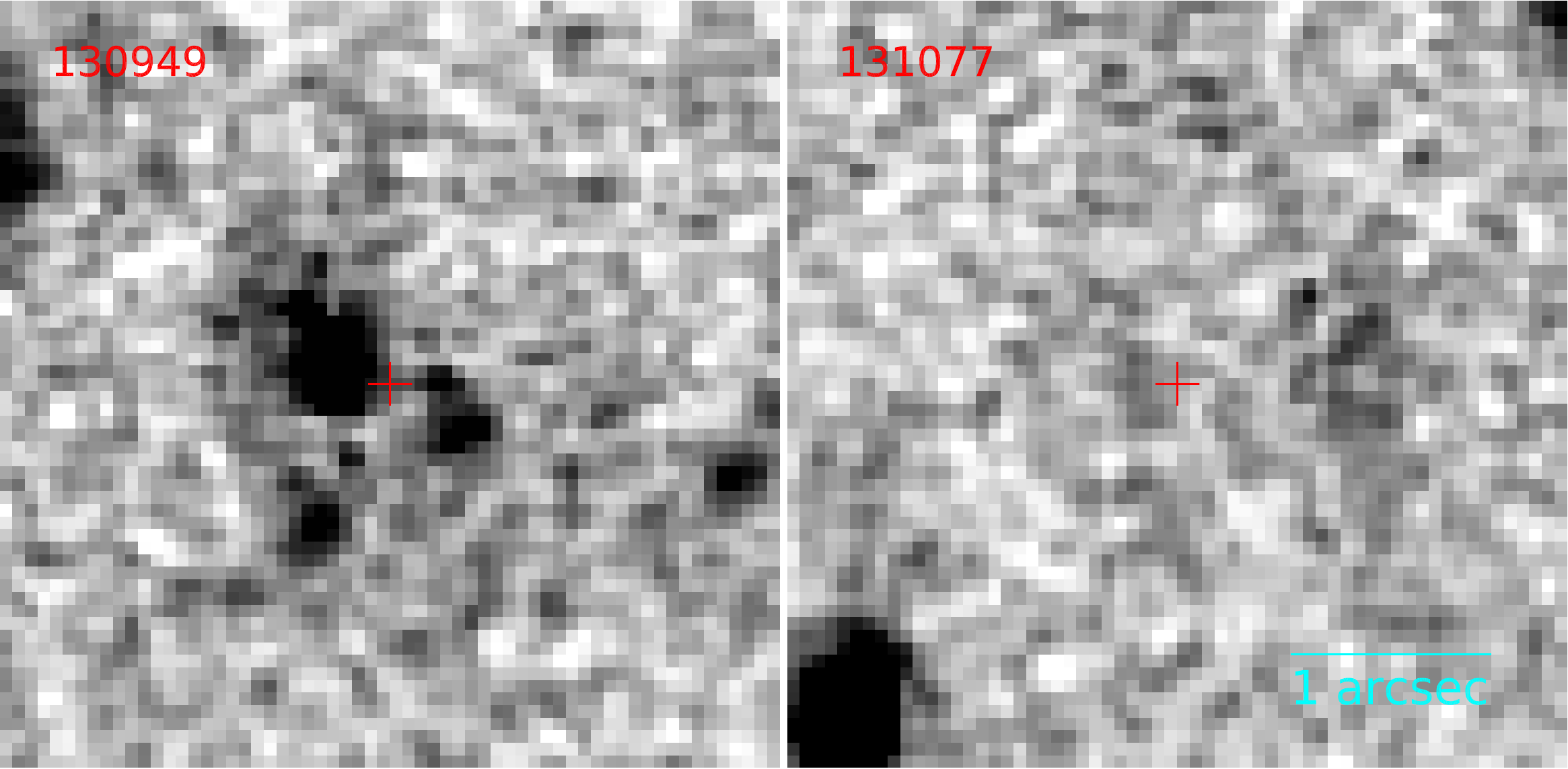}
\caption{HST/WFC3 $J_{110}$-band stamp images of the two massive starbursts in the cluster core. The crosses mark the source positions in the UltraVista $K_{s}$-band. 
\label{Fig:sb_stamp}}
\end{figure}

Despite the presence of a cluster-like environment (including both a collapsed massive halo and a high concentration of massive galaxies in the core), the fraction of galaxies that are classified as quiescent in CL J1001 at $M_{*} > 10^{11} M_{\odot}$ is estimated to be less than $\sim20\%$. This quiescent fraction is similar to that in the field and is significantly lower than that in known $z\sim 2$ mature clusters, suggesting that most central cluster galaxies will be quenched only after they accrete onto the cluster.  
This is different from the ``pre-processing" scenario in which galaxies are quenched in groups or large-scale filaments prior to cluster assembly, due to, e.g., strangulation. We speculate that this might be due to the fact that at high redshifts, only a small fraction of cluster galaxies were accreted onto the final cluster halo as a member of a group-sized halo with $M_{200c} > 10^{13} M_{\odot}$, as suggested by simulations~\citep{Balogh:2009,McGee:2009}. Moreover, simulations suggest that cold streams can penetrate moderately massive, group-sized halos with $M_{200c} \sim$10$^{13} M_{\odot}$ at $z \gtrsim 2$~\citep{Dekel:2006}, which makes it difficult to fully quench protocluster galaxies even if they were located in a group-sized halo before their accretion onto the cluster. These arguments are also consistent with the fact that is are no clear evidence for suppressed star formation or an elevated quiescent fraction in protocluster regions.  In fact, based on current studies of $z \gtrsim 2$ protoclusters, star formation in protoclusters seems to be elevated~\citep{Casey:2016} rather than suppressed compared to field galaxies.

We caution that while many studies of high-redshift protoclusters observed a high volume density of highly SFGs or dusty galaxies (detected down to some flux level in the far-infrared or submillimeter), it is unclear whether this is caused by elevated SFRs in individual galaxies or simply caused by the fact that there are more massive galaxies in protocluster regions. 
A thorough examination of the mass-star formation relation as well as the relative fraction of quiescent and SFGs in protocluster regions, and its comparison to field galaxies, is required to obtain solid conclusions. This is, however,  quite difficult for protoclusters due to their extended region (several tens of Mpc) and lower significance of galaxy overdensities, which inhibit a census of its member galaxies (particularly quiescent ones). Moreover, simulations suggest that a significant fraction of protocluster galaxies may not end up in clusters at $z \sim 0$, especially those low-mass galaxies in the outskirts~\citep{Contini:2016}. This makes it more complicated to compare observations of protoclusters and simulations.

Although the current observed quiescent fraction in CL J1001 is similar to that in the field, several pieces of evidence suggest that this fraction will increase over a short time scale. The total molecular gas mass for galaxies in the core is estimated to be $M_{gas} \sim 5 \times 10^{11} M_{\odot}$ (Section~\ref{Sec:galaxies}). Considering their current SFR, this structure will consume all the available gas within $\sim150-200$ Myr (twice larger if using Chabrier IMF for the stellar mass estimation). The halo mass of this structure ($\sim10^{14} M_{\odot}$) falls in the regime where the infalling gas is fully shock-heated instead of forming cold streams~\citep{Dekel:2006}. These properties suggest that the cluster will likely form a predominant population of quiescent galaxies in the core by $z \sim 2.2$. 
Moreover, most of the massive SFGs in the core already fall on the mass-size relation of quiescent galaxies at the same redshift (Fig.~\ref{Fig:mass_size}), providing further evidence that these galaxies may soon transform into quiescent galaxies.

Compared to field galaxies, massive galaxies in the core of CL J1001 exhibit a higher starburst fraction, suggesting that cluster ellipticals may form their stars through more violent starbursting events and in shorter time scales than field galaxies. This is consistent with galaxy cluster archeology studies~\citep{Thomas:2005}. This high starburst fraction may be partly due to a higher merger rate, as expected for this high density region with moderate velocity dispersion. Indeed, one of the starbursting galaxies (ID 130949) appears to be a complex multi-component galaxy system in the high-resolution WFC3 $J_{110}$ image while the other one (ID 131077) is  completely undetected due to obscuration and also the shallow depth of $J_{110}$ imaging~(Fig.~\ref{Fig:sb_stamp}). The high starburst fraction could be also caused by the compression on the molecular gas in galaxies by the hot intracluster medium (IGM), which could trigger the collapse of molecular clouds and lead to efficient star formation within a short timescale~\citep{Fujita:1999,Bekki:2003}. Deeper WFC3 imaging in the rest-frame optical and spatially resolved distribution of both star formation and molecular gas are required to provide further insights into these questions, which we defer to a future work.

\section{Conclusion}
\label{Sec:conclusion}

We conclude that we have found a massive galaxy overdensity at $z = 2.506$, which is likely to be the most distant X-ray cluster known to date. This overdensity is embedded in a collapsed, cluster-sized halo as suggested by its high density of massive galaxies in the core, extended X-ray emission, cluster-like mass density profile, and velocity dispersion of member galaxies. Moreover, this structure exhibits both a high SFR density and a predominant population of massive SFGs in the core, indicating that it is in a major formation phase for the central massive cluster galaxies when most of them have not been quenched. These properties differentiate this structure from other structures recently discovered at similar redshifts and suggest that it may represent the missing link between mature clusters and protoclusters. The following are our findings on its main properties and its implications for cluster formation.

1. The structure was identified as the most significant overdensity of DRGs in COSMOS with 11 DRGs distributed over a 80kpc region. It is also the brightest \textit{Herschel}/SPIRE source (unresolved) in the central COSMOS region covered by CANDELS-\textit{Herschel} survey with flux densities 60-80 mJy in the SPIRE bands. 

2. Extensive follow-up observations with IRAM-NOEMA, VLT/KMOS, and JVLA spectroscopically confirmed 17 members including 7 DRGs in the core. Based on these 17 members, the cluster redshift is determined to be $z = 2.506$ with a velocity dispersion of $530\pm120$ km s$^{-1}$.

3. Combining \textit{XMM-Newton} and \textit{Chandra} observations of the field, the overdensity exhibits extended X-ray emission at the 4$\sigma$ confidence level with an X-ray luminosity $L_{0.1-2.4 \mathrm{keV}} = 8.8 \times 10^{43}$ erg s$^{-1}$. The X-ray luminosity,  the velocity dispersion, and the stellar mass content of this structure all suggest a total halo mass of $M_{200c} = 10^{13.9\pm0.2} M_{\odot}$.

4. The structure exhibits a high star formation density in the 80kpc core with a combined SFR $\sim$3400 $M_{\odot}$ yr$^{-1}$ and a gas depletion time of $\sim$200 Myr. Galaxies in the core show both elevated starburst activities and supermassive black hole accretion compared to field galaxies. 

5. The core of this structure is dominated by SFGs with only 2 out the 11 DRGs classified as quiescent galaxies. At $M_{*} > 10^{11} M_{\odot}$ the quiescent fraction is around $\sim$20\% without significant dependence on the clustercentric distance up to $\sim$ 700 kpc from the core. This quiescent fraction is similar to field galaxies at the same redshift and significantly lower than that in previously discovered mature clusters at $z \sim 2$.

6. At the massive end ($M_{*} > 10^{11} M_{\odot}$), both quiescent galaxies and star-forming ones in the core of the structure appear to be compact, which is consistent with the mass-size relation for quiescent galaxies in the field.

One of the most prominent features of this structure is the presence of both a collapsed, cluster-sized halo and a high abundance of massive, highly SFGs. Its discovery suggests that most cluster ellipticals likely formed only after their accretion onto a cluster-sized halo, though more similar structures are needed to confirm. Recently a number of Planck sources with high \textit{Herschel}/SPIRE fluxes have been discovered and are likely to be $z \gtrsim 2$ (proto)cluster candidates~\citep{PlanckXIII:2015}, some of which may be in massive halos similar to this structure, although further follow-up observations are needed for detailed comparisons.  
Future studies of a large number of similar structures will definitively clarify the formation path of massive galaxy clusters and provide critical constraints on cosmology.
 
\begin{acknowledgements}

We thank the anonymous referee for helpful comments which improved the content of this paper. 
This study is based on observations with the IRAM Plateau de Bure Interferometer. IRAM is supported by INSU/CNRS (France), MPG (Germany), and IGN (Spain).  
This paper makes use of the data from the Karl G. Jansky Very Large Array. The National Radio Astronomy Observatory is a facility of the National Science Foundation operated under cooperative agreement by Associated Universities, Inc.
This paper makes use of the following ALMA data: ADS/JAO.ALMA\#2011.0.00539.S. ALMA is a partnership of ESO (representing its member states), NSF (USA), and NINS (Japan), together with NRC (Canada), NSC, ASIAA (Taiwan), and KASI (Republic of Korea), in cooperation with the Republic of Chile. The Joint ALMA Observatory is operated by ESO, AUI/NRAO, and NAOJ. 
This work is partly based on observations collected at the European Organization for Astronomical Research in the Southern Hemisphere under ESO program 096.A-0891(A).

This research is supported by the European Commission through the FP7 SPACE project ASTRODEEP (Ref.No: 312725). T.W. acknowledges support for this work from the National Natural Science Foundation of China under grant 11303014.

Facilities: HST, $Herschel$(PACS, SPIRE), $Spitzer$ (IRAC, MIPS), $Chandra$, $XMM-Newton$.
\end{acknowledgements}

\bibliographystyle{aasjournal}
\bibliography{ms.bbl}

\begin{thebibliography}{}
\expandafter\ifx\csname natexlab\endcsname\relax\def\natexlab#1{#1}\fi

\bibitem[{{Allevato} {et~al.}(2012){Allevato}, {Finoguenov}, {Hasinger},
  {Miyaji}, {Cappelluti}, {Salvato}, {Zamorani}, {Gilli}, {George}, {Tanaka},
  {Brusa}, {Silverman}, {Civano}, {Elvis}, \& {Shankar}}]{Allevato:2012}
{Allevato}, V., {Finoguenov}, A., {Hasinger}, G., {et~al.} 2012, \apj, 758, 47

\bibitem[{{Andreon}(1996)}]{Andreon:1996}
{Andreon}, S. 1996, \aap, 314, 763

\bibitem[{{Andreon} {et~al.}(2014){Andreon}, {Newman}, {Trinchieri},
  {Raichoor}, {Ellis}, \& {Treu}}]{Andreon:2014}
{Andreon}, S., {Newman}, A.~B., {Trinchieri}, G., {et~al.} 2014, \aap, 565,
  A120

\bibitem[{{Aretxaga} {et~al.}(2011){Aretxaga}, {Wilson}, {Aguilar}, {Alberts},
  {Scott}, {Scoville}, {Yun}, {Austermann}, {Downes}, {Ezawa}, {Hatsukade},
  {Hughes}, {Kawabe}, {Kohno}, {Oshima}, {Perera}, {Tamura}, \&
  {Zeballos}}]{Aretxaga:2011}
{Aretxaga}, I., {Wilson}, G.~W., {Aguilar}, E., {et~al.} 2011, \mnras, 415,
  3831

\bibitem[{{Balogh} {et~al.}(2009){Balogh}, {McGee}, {Wilman}, {Bower}, {Hau},
  {Morris}, {Mulchaey}, {Oemler}, {Parker}, \& {Gwyn}}]{Balogh:2009}
{Balogh}, M.~L., {McGee}, S.~L., {Wilman}, D., {et~al.} 2009, \mnras, 398, 754

\bibitem[{{Bassett} {et~al.}(2013){Bassett}, {Papovich}, {Lotz}, {Bell},
  {Finkelstein}, {Newman}, {Tran}, {Almaini}, {Lani}, {Cooper}, {Croton},
  {Dekel}, {Ferguson}, {Kocevski}, {Koekemoer}, {Koo}, {McGrath}, {McIntosh},
  \& {Wechsler}}]{Bassett:2013}
{Bassett}, R., {Papovich}, C., {Lotz}, J.~M., {et~al.} 2013, \apj, 770, 58

\bibitem[{{Beers} {et~al.}(1990){Beers}, {Flynn}, \& {Gebhardt}}]{Beers:1990}
{Beers}, T.~C., {Flynn}, K., \& {Gebhardt}, K. 1990, \aj, 100, 32

\bibitem[{{Behroozi} {et~al.}(2013){Behroozi}, {Wechsler}, \&
  {Conroy}}]{Behroozi:2013a}
{Behroozi}, P.~S., {Wechsler}, R.~H., \& {Conroy}, C. 2013, \apj, 770, 57

\bibitem[{{Bekki} \& {Couch}(2003)}]{Bekki:2003}
{Bekki}, K., \& {Couch}, W.~J. 2003, \apjl, 596, L13

\bibitem[{{Berrier} {et~al.}(2009){Berrier}, {Stewart}, {Bullock}, {Purcell},
  {Barton}, \& {Wechsler}}]{Berrier:2009}
{Berrier}, J.~C., {Stewart}, K.~R., {Bullock}, J.~S., {et~al.} 2009, \apj, 690,
  1292

\bibitem[{{B{\'e}thermin} {et~al.}(2010){B{\'e}thermin}, {Dole}, {Beelen}, \&
  {Aussel}}]{Bethermin:2010}
{B{\'e}thermin}, M., {Dole}, H., {Beelen}, A., \& {Aussel}, H. 2010, \aap, 512,
  A78

\bibitem[{{B{\'e}thermin} {et~al.}(2014){B{\'e}thermin}, {Kilbinger}, {Daddi},
  {Gabor}, {Finoguenov}, {McCracken}, {Wolk}, {Aussel}, {Strazzulo}, {Le
  Floc'h}, {Gobat}, {Rodighiero}, {Dickinson}, {Wang}, {Lutz}, \&
  {Heinis}}]{Bethermin:2014}
{B{\'e}thermin}, M., {Kilbinger}, M., {Daddi}, E., {et~al.} 2014, \aap, 567,
  A103

\bibitem[{{Boylan-Kolchin} {et~al.}(2009){Boylan-Kolchin}, {Springel}, {White},
  {Jenkins}, \& {Lemson}}]{Boylan-Kolchin:2009}
{Boylan-Kolchin}, M., {Springel}, V., {White}, S.~D.~M., {Jenkins}, A., \&
  {Lemson}, G. 2009, \mnras, 398, 1150

\bibitem[{{Brammer} {et~al.}(2008){Brammer}, {van Dokkum}, \&
  {Coppi}}]{Brammer:2008}
{Brammer}, G.~B., {van Dokkum}, P.~G., \& {Coppi}, P. 2008, \apj, 686, 1503

\bibitem[{{Brammer} {et~al.}(2012){Brammer}, {van Dokkum}, {Franx},
  {Fumagalli}, {Patel}, {Rix}, {Skelton}, {Kriek}, {Nelson}, {Schmidt},
  {Bezanson}, {da Cunha}, {Erb}, {Fan}, {F{\"o}rster Schreiber}, {Illingworth},
  {Labb{\'e}}, {Leja}, {Lundgren}, {Magee}, {Marchesini}, {McCarthy},
  {Momcheva}, {Muzzin}, {Quadri}, {Steidel}, {Tal}, {Wake}, {Whitaker}, \&
  {Williams}}]{Brammer:2012}
{Brammer}, G.~B., {van Dokkum}, P.~G., {Franx}, M., {et~al.} 2012, \apjs, 200,
  13

\bibitem[{{Brodwin} {et~al.}(2013){Brodwin}, {Stanford}, {Gonzalez}, {Zeimann},
  {Snyder}, {Mancone}, {Pope}, {Eisenhardt}, {Stern}, {Alberts}, {Ashby},
  {Brown}, {Chary}, {Dey}, {Galametz}, {Gettings}, {Jannuzi}, {Miller},
  {Moustakas}, \& {Moustakas}}]{Brodwin:2013}
{Brodwin}, M., {Stanford}, S.~A., {Gonzalez}, A.~H., {et~al.} 2013, \apj, 779,
  138

\bibitem[{{Bruzual} \& {Charlot}(2003)}]{Bruzual:2003}
{Bruzual}, G., \& {Charlot}, S. 2003, \mnras, 344, 1000

\bibitem[{{Bussmann} {et~al.}(2015){Bussmann}, {Riechers}, {Fialkov},
  {Scudder}, {Hayward}, {Cowley}, {Bock}, {Calanog}, {Chapman}, {Cooray}, {De
  Bernardis}, {Farrah}, {Fu}, {Gavazzi}, {Hopwood}, {Ivison}, {Jarvis},
  {Lacey}, {Loeb}, {Oliver}, {P{\'e}rez-Fournon}, {Rigopoulou}, {Roseboom},
  {Scott}, {Smith}, {Vieira}, {Wang}, \& {Wardlow}}]{Bussmann:2015}
{Bussmann}, R.~S., {Riechers}, D., {Fialkov}, A., {et~al.} 2015, \apj, 812, 43

\bibitem[{{Calzetti} {et~al.}(2000){Calzetti}, {Armus}, {Bohlin}, {Kinney},
  {Koornneef}, \& {Storchi-Bergmann}}]{Calzetti:2000}
{Calzetti}, D., {Armus}, L., {Bohlin}, R.~C., {et~al.} 2000, \apj, 533, 682

\bibitem[{{Capak} {et~al.}(2011){Capak}, {Riechers}, {Scoville}, {Carilli},
  {Cox}, {Neri}, {Robertson}, {Salvato}, {Schinnerer}, {Yan}, {Wilson}, {Yun},
  {Civano}, {Elvis}, {Karim}, {Mobasher}, \& {Staguhn}}]{Capak:2011}
{Capak}, P.~L., {Riechers}, D., {Scoville}, N.~Z., {et~al.} 2011, \nat, 470,
  233

\bibitem[{{Cappelluti} {et~al.}(2009){Cappelluti}, {Brusa}, {Hasinger},
  {Comastri}, {Zamorani}, {Finoguenov}, {Gilli}, {Puccetti}, {Miyaji},
  {Salvato}, {Vignali}, {Aldcroft}, {B{\"o}hringer}, {Brunner}, {Civano},
  {Elvis}, {Fiore}, {Fruscione}, {Griffiths}, {Guzzo}, {Iovino}, {Koekemoer},
  {Mainieri}, {Scoville}, {Shopbell}, {Silverman}, \& {Urry}}]{Cappelluti:2009}
{Cappelluti}, N., {Brusa}, M., {Hasinger}, G., {et~al.} 2009, \aap, 497, 635

\bibitem[{{Cappelluti} {et~al.}(2013){Cappelluti}, {Kashlinsky}, {Arendt},
  {Comastri}, {Fazio}, {Finoguenov}, {Hasinger}, {Mather}, {Miyaji}, \&
  {Moseley}}]{Cappelluti:2013}
{Cappelluti}, N., {Kashlinsky}, A., {Arendt}, R.~G., {et~al.} 2013, \apj, 769,
  68

\bibitem[{{Casey}(2012)}]{Casey:2012a}
{Casey}, C.~M. 2012, \mnras, 425, 3094

\bibitem[{{Casey}(2016)}]{Casey:2016}
---. 2016, \apj, 824, 36

\bibitem[{{Casey} {et~al.}(2013){Casey}, {Chen}, {Cowie}, {Barger}, {Capak},
  {Ilbert}, {Koss}, {Lee}, {Le Floc'h}, {Sanders}, \& {Williams}}]{Casey:2013}
{Casey}, C.~M., {Chen}, C.-C., {Cowie}, L.~L., {et~al.} 2013, \mnras, 436, 1919

\bibitem[{{Casey} {et~al.}(2015){Casey}, {Cooray}, {Capak}, {Fu}, {Kovac},
  {Lilly}, {Sanders}, {Scoville}, \& {Treister}}]{Casey:2015}
{Casey}, C.~M., {Cooray}, A., {Capak}, P., {et~al.} 2015, \apjl, 808, L33

\bibitem[{{Chabrier}(2003)}]{Chabrier:2003}
{Chabrier}, G. 2003, \pasp, 115, 763

\bibitem[{{Chapman} {et~al.}(2009){Chapman}, {Blain}, {Ibata}, {Ivison},
  {Smail}, \& {Morrison}}]{Chapman:2009}
{Chapman}, S.~C., {Blain}, A., {Ibata}, R., {et~al.} 2009, \apj, 691, 560

\bibitem[{{Chary} \& {Elbaz}(2001)}]{Chary:2001}
{Chary}, R., \& {Elbaz}, D. 2001, \apj, 556, 562

\bibitem[{{Chiang} {et~al.}(2013){Chiang}, {Overzier}, \&
  {Gebhardt}}]{Chiang:2013}
{Chiang}, Y.-K., {Overzier}, R., \& {Gebhardt}, K. 2013, \apj, 779, 127

\bibitem[{{Chiang} {et~al.}(2014){Chiang}, {Overzier}, \&
  {Gebhardt}}]{Chiang:2014}
---. 2014, \apjl, 782, L3

\bibitem[{{Civano} {et~al.}(2016){Civano}, {Marchesi}, {Comastri}, {Urry},
  {Elvis}, {Cappelluti}, {Puccetti}, {Brusa}, {Zamorani}, {Hasinger},
  {Aldcroft}, {Alexander}, {Allevato}, {Brunner}, {Capak}, {Finoguenov},
  {Fiore}, {Fruscione}, {Gilli}, {Glotfelty}, {Griffiths}, {Hao}, {Harrison},
  {Jahnke}, {Kartaltepe}, {Karim}, {LaMassa}, {Lanzuisi}, {Miyaji}, {Ranalli},
  {Salvato}, {Sargent}, {Scoville}, {Schawinski}, {Schinnerer}, {Silverman},
  {Smolcic}, {Stern}, {Toft}, {Trakhtenbrot}, {Treister}, \&
  {Vignali}}]{Civano:2016}
{Civano}, F., {Marchesi}, S., {Comastri}, A., {et~al.} 2016, \apj, 819, 62

\bibitem[{{Clements} {et~al.}(2014){Clements}, {Braglia}, {Hyde},
  {P{\'e}rez-Fournon}, {Bock}, {Cava}, {Chapman}, {Conley}, {Cooray}, {Farrah},
  {Gonz{\'a}lez Solares}, {Marchetti}, {Marsden}, {Oliver}, {Roseboom},
  {Schulz}, {Smith}, {Vaccari}, {Vieira}, {Viero}, {Wang}, {Wardlow}, {Zemcov},
  \& {de Zotti}}]{Clements:2014}
{Clements}, D.~L., {Braglia}, F.~G., {Hyde}, A.~K., {et~al.} 2014, \mnras, 439,
  1193

\bibitem[{{Contini} {et~al.}(2016){Contini}, {De Lucia}, {Hatch}, {Borgani}, \&
  {Kang}}]{Contini:2016}
{Contini}, E., {De Lucia}, G., {Hatch}, N., {Borgani}, S., \& {Kang}, X. 2016,
  \mnras, 456, 1924

\bibitem[{{Daddi} {et~al.}(2009){Daddi}, {Dannerbauer}, {Stern}, {Dickinson},
  {Morrison}, {Elbaz}, {Giavalisco}, {Mancini}, {Pope}, \&
  {Spinrad}}]{Daddi:2009a}
{Daddi}, E., {Dannerbauer}, H., {Stern}, D., {et~al.} 2009, \apj, 694, 1517

\bibitem[{{Danese} {et~al.}(1980){Danese}, {de Zotti}, \& {di
  Tullio}}]{Danese:1980}
{Danese}, L., {de Zotti}, G., \& {di Tullio}, G. 1980, \aap, 82, 322

\bibitem[{{De Lucia} \& {Blaizot}(2007)}]{DeLucia:2007}
{De Lucia}, G., \& {Blaizot}, J. 2007, \mnras, 375, 2

\bibitem[{{De Lucia} {et~al.}(2012){De Lucia}, {Weinmann}, {Poggianti},
  {Arag{\'o}n-Salamanca}, \& {Zaritsky}}]{DeLucia:2012}
{De Lucia}, G., {Weinmann}, S., {Poggianti}, B.~M., {Arag{\'o}n-Salamanca}, A.,
  \& {Zaritsky}, D. 2012, \mnras, 423, 1277

\bibitem[{{Dekel} \& {Birnboim}(2006)}]{Dekel:2006}
{Dekel}, A., \& {Birnboim}, Y. 2006, \mnras, 368, 2

\bibitem[{{Delaye} {et~al.}(2014){Delaye}, {Huertas-Company}, {Mei}, {Lidman},
  {Licitra}, {Newman}, {Raichoor}, {Shankar}, {Barrientos}, {Bernardi},
  {Cerulo}, {Couch}, {Demarco}, {Mu{\~n}oz}, {S{\'a}nchez-Janssen}, \&
  {Tanaka}}]{Delaye:2014}
{Delaye}, L., {Huertas-Company}, M., {Mei}, S., {et~al.} 2014, \mnras, 441, 203

\bibitem[{{Diener} {et~al.}(2013){Diener}, {Lilly}, {Knobel}, {Zamorani},
  {Lemson}, {Kampczyk}, {Scoville}, {Carollo}, {Contini}, {Kneib}, {Le Fevre},
  {Mainieri}, {Renzini}, {Scodeggio}, {Bardelli}, {Bolzonella}, {Bongiorno},
  {Caputi}, {Cucciati}, {de la Torre}, {de Ravel}, {Franzetti}, {Garilli},
  {Iovino}, {Kova{\v c}}, {Lamareille}, {Le Borgne}, {Le Brun}, {Maier},
  {Mignoli}, {Pello}, {Peng}, {Perez Montero}, {Presotto}, {Silverman},
  {Tanaka}, {Tasca}, {Tresse}, {Vergani}, {Zucca}, {Bordoloi}, {Cappi},
  {Cimatti}, {Coppa}, {Koekemoer}, {L{\'o}pez-Sanjuan}, {McCracken}, {Moresco},
  {Nair}, {Pozzetti}, \& {Welikala}}]{Diener:2013}
{Diener}, C., {Lilly}, S.~J., {Knobel}, C., {et~al.} 2013, \apj, 765, 109

\bibitem[{{Diener} {et~al.}(2015){Diener}, {Lilly}, {Ledoux}, {Zamorani},
  {Bolzonella}, {Murphy}, {Capak}, {Ilbert}, \& {McCracken}}]{Diener:2015}
{Diener}, C., {Lilly}, S.~J., {Ledoux}, C., {et~al.} 2015, \apj, 802, 31

\bibitem[{{Elbaz} {et~al.}(2007){Elbaz}, {Daddi}, {Le Borgne}, {Dickinson},
  {Alexander}, {Chary}, {Starck}, {Brandt}, {Kitzbichler}, {MacDonald},
  {Nonino}, {Popesso}, {Stern}, \& {Vanzella}}]{Elbaz:2007}
{Elbaz}, D., {Daddi}, E., {Le Borgne}, D., {et~al.} 2007, \aap, 468, 33

\bibitem[{{Elbaz} {et~al.}(2011){Elbaz}, {Dickinson}, {Hwang},
  {D{\'{\i}}az-Santos}, {Magdis}, {Magnelli}, {Le Borgne}, {Galliano},
  {Pannella}, {Chanial}, {Armus}, {Charmandaris}, {Daddi}, {Aussel}, {Popesso},
  {Kartaltepe}, {Altieri}, {Valtchanov}, {Coia}, {Dannerbauer}, {Dasyra},
  {Leiton}, {Mazzarella}, {Alexander}, {Buat}, {Burgarella}, {Chary}, {Gilli},
  {Ivison}, {Juneau}, {Le Floc'h}, {Lutz}, {Morrison}, {Mullaney}, {Murphy},
  {Pope}, {Scott}, {Brodwin}, {Calzetti}, {Cesarsky}, {Charlot}, {Dole},
  {Eisenhardt}, {Ferguson}, {F{\"o}rster Schreiber}, {Frayer}, {Giavalisco},
  {Huynh}, {Koekemoer}, {Papovich}, {Reddy}, {Surace}, {Teplitz}, {Yun}, \&
  {Wilson}}]{Elbaz:2011}
{Elbaz}, D., {Dickinson}, M., {Hwang}, H.~S., {et~al.} 2011, \aap, 533, A119

\bibitem[{{Elvis} {et~al.}(2009){Elvis}, {Civano}, {Vignali}, {Puccetti},
  {Fiore}, {Cappelluti}, {Aldcroft}, {Fruscione}, {Zamorani}, {Comastri},
  {Brusa}, {Gilli}, {Miyaji}, {Damiani}, {Koekemoer}, {Finoguenov}, {Brunner},
  {Urry}, {Silverman}, {Mainieri}, {Hasinger}, {Griffiths}, {Carollo}, {Hao},
  {Guzzo}, {Blain}, {Calzetti}, {Carilli}, {Capak}, {Ettori}, {Fabbiano},
  {Impey}, {Lilly}, {Mobasher}, {Rich}, {Salvato}, {Sanders}, {Schinnerer},
  {Scoville}, {Shopbell}, {Taylor}, {Taniguchi}, \& {Volonteri}}]{Elvis:2009}
{Elvis}, M., {Civano}, F., {Vignali}, C., {et~al.} 2009, \apjs, 184, 158

\bibitem[{{Erb} {et~al.}(2006){Erb}, {Shapley}, {Pettini}, {Steidel}, {Reddy},
  \& {Adelberger}}]{Erb:2006}
{Erb}, D.~K., {Shapley}, A.~E., {Pettini}, M., {et~al.} 2006, \apj, 644, 813

\bibitem[{{Erfanianfar} {et~al.}(2013){Erfanianfar}, {Finoguenov}, {Tanaka},
  {Lerchster}, {Nandra}, {Laird}, {Connelly}, {Bielby}, {Mirkazemi}, {Faber},
  {Kocevski}, {Cooper}, {Newman}, {Jeltema}, {Coil}, {Brimioulle}, {Davis},
  {McCracken}, {Willmer}, {Gerke}, {Cappelluti}, \& {Gwyn}}]{Erfanianfar:2013}
{Erfanianfar}, G., {Finoguenov}, A., {Tanaka}, M., {et~al.} 2013, \apj, 765,
  117

\bibitem[{{Evrard} {et~al.}(2008){Evrard}, {Bialek}, {Busha}, {White}, {Habib},
  {Heitmann}, {Warren}, {Rasia}, {Tormen}, {Moscardini}, {Power}, {Jenkins},
  {Gao}, {Frenk}, {Springel}, {White}, \& {Diemand}}]{Evrard:2008}
{Evrard}, A.~E., {Bialek}, J., {Busha}, M., {et~al.} 2008, \apj, 672, 122

\bibitem[{{Fabian} {et~al.}(2003){Fabian}, {Sanders}, {Crawford}, \&
  {Ettori}}]{Fabian:2003}
{Fabian}, A.~C., {Sanders}, J.~S., {Crawford}, C.~S., \& {Ettori}, S. 2003,
  \mnras, 341, 729

\bibitem[{{Finoguenov} {et~al.}(2007){Finoguenov}, {Guzzo}, {Hasinger},
  {Scoville}, {Aussel}, {B{\"o}hringer}, {Brusa}, {Capak}, {Cappelluti},
  {Comastri}, {Giodini}, {Griffiths}, {Impey}, {Koekemoer}, {Kneib},
  {Leauthaud}, {Le F{\`e}vre}, {Lilly}, {Mainieri}, {Massey}, {McCracken},
  {Mobasher}, {Murayama}, {Peacock}, {Sakelliou}, {Schinnerer}, {Silverman},
  {Smol{\v c}i{\'c}}, {Taniguchi}, {Tasca}, {Taylor}, {Trump}, \&
  {Zamorani}}]{Finoguenov:2007}
{Finoguenov}, A., {Guzzo}, L., {Hasinger}, G., {et~al.} 2007, \apjs, 172, 182

\bibitem[{{Finoguenov} {et~al.}(2009){Finoguenov}, {Connelly}, {Parker},
  {Wilman}, {Mulchaey}, {Saglia}, {Balogh}, {Bower}, \&
  {McGee}}]{Finoguenov:2009}
{Finoguenov}, A., {Connelly}, J.~L., {Parker}, L.~C., {et~al.} 2009, \apj, 704,
  564

\bibitem[{{Finoguenov} {et~al.}(2015){Finoguenov}, {Tanaka}, {Cooper},
  {Allevato}, {Cappelluti}, {Choi}, {Heymans}, {Bauer}, {Ziparo}, {Ranalli},
  {Silverman}, {Brandt}, {Xue}, {Mulchaey}, {Howes}, {Schmid}, {Wilman},
  {Comastri}, {Hasinger}, {Mainieri}, {Luo}, {Tozzi}, {Rosati}, {Capak}, \&
  {Popesso}}]{Finoguenov:2015}
{Finoguenov}, A., {Tanaka}, M., {Cooper}, M., {et~al.} 2015, \aap, 576, A130

\bibitem[{{Franx} {et~al.}(2003){Franx}, {Labb{\'e}}, {Rudnick}, {van Dokkum},
  {Daddi}, {F{\"o}rster Schreiber}, {Moorwood}, {Rix}, {R{\"o}ttgering}, {van
  de Wel}, {van der Werf}, \& {van Starkenburg}}]{Franx:2003}
{Franx}, M., {Labb{\'e}}, I., {Rudnick}, G., {et~al.} 2003, \apjl, 587, L79

\bibitem[{{Fujita} \& {Nagashima}(1999)}]{Fujita:1999}
{Fujita}, Y., \& {Nagashima}, M. 1999, \apj, 516, 619

\bibitem[{{Gobat} {et~al.}(2011){Gobat}, {Daddi}, {Onodera}, {Finoguenov},
  {Renzini}, {Arimoto}, {Bouwens}, {Brusa}, {Chary}, {Cimatti}, {Dickinson},
  {Kong}, \& {Mignoli}}]{Gobat:2011}
{Gobat}, R., {Daddi}, E., {Onodera}, M., {et~al.} 2011, \aap, 526, A133

\bibitem[{{Gobat} {et~al.}(2013){Gobat}, {Strazzullo}, {Daddi}, {Onodera},
  {Carollo}, {Renzini}, {Finoguenov}, {Cimatti}, {Scarlata}, \&
  {Arimoto}}]{Gobat:2013}
{Gobat}, R., {Strazzullo}, V., {Daddi}, E., {et~al.} 2013, \apj, 776, 9

\bibitem[{{Granato} {et~al.}(2015){Granato}, {Ragone-Figueroa},
  {Dom{\'{\i}}nguez-Tenreiro}, {Obreja}, {Borgani}, {De Lucia}, \&
  {Murante}}]{Granato:2015}
{Granato}, G.~L., {Ragone-Figueroa}, C., {Dom{\'{\i}}nguez-Tenreiro}, R.,
  {et~al.} 2015, \mnras, 450, 1320

\bibitem[{{Grogin} {et~al.}(2011){Grogin}, {Kocevski}, {Faber}, {Ferguson},
  {Koekemoer}, {Riess}, {Acquaviva}, {Alexander}, {Almaini}, {Ashby}, {Barden},
  {Bell}, {Bournaud}, {Brown}, {Caputi}, {Casertano}, {Cassata}, {Castellano},
  {Challis}, {Chary}, {Cheung}, {Cirasuolo}, {Conselice}, {Roshan Cooray},
  {Croton}, {Daddi}, {Dahlen}, {Dav{\'e}}, {de Mello}, {Dekel}, {Dickinson},
  {Dolch}, {Donley}, {Dunlop}, {Dutton}, {Elbaz}, {Fazio}, {Filippenko},
  {Finkelstein}, {Fontana}, {Gardner}, {Garnavich}, {Gawiser}, {Giavalisco},
  {Grazian}, {Guo}, {Hathi}, {H{\"a}ussler}, {Hopkins}, {Huang}, {Huang},
  {Jha}, {Kartaltepe}, {Kirshner}, {Koo}, {Lai}, {Lee}, {Li}, {Lotz}, {Lucas},
  {Madau}, {McCarthy}, {McGrath}, {McIntosh}, {McLure}, {Mobasher},
  {Moustakas}, {Mozena}, {Nandra}, {Newman}, {Niemi}, {Noeske}, {Papovich},
  {Pentericci}, {Pope}, {Primack}, {Rajan}, {Ravindranath}, {Reddy}, {Renzini},
  {Rix}, {Robaina}, {Rodney}, {Rosario}, {Rosati}, {Salimbeni}, {Scarlata},
  {Siana}, {Simard}, {Smidt}, {Somerville}, {Spinrad}, {Straughn}, {Strolger},
  {Telford}, {Teplitz}, {Trump}, {van der Wel}, {Villforth}, {Wechsler},
  {Weiner}, {Wiklind}, {Wild}, {Wilson}, {Wuyts}, {Yan}, \&
  {Yun}}]{Grogin:2011}
{Grogin}, N.~A., {Kocevski}, D.~D., {Faber}, S.~M., {et~al.} 2011, \apjs, 197,
  35

\bibitem[{{Gunn} \& {Gott}(1972)}]{Gunn:1972}
{Gunn}, J.~E., \& {Gott}, III, J.~R. 1972, \apj, 176, 1

\bibitem[{{Guo} {et~al.}(2011){Guo}, {White}, {Boylan-Kolchin}, {De Lucia},
  {Kauffmann}, {Lemson}, {Li}, {Springel}, \& {Weinmann}}]{GuoQ:2011}
{Guo}, Q., {White}, S., {Boylan-Kolchin}, M., {et~al.} 2011, \mnras, 413, 101

\bibitem[{{Harris} \& {Grindlay}(1979)}]{Harris:1979}
{Harris}, D.~E., \& {Grindlay}, J.~E. 1979, \mnras, 188, 25

\bibitem[{{Hayashi} {et~al.}(2012){Hayashi}, {Kodama}, {Tadaki}, {Koyama}, \&
  {Tanaka}}]{Hayashi:2012}
{Hayashi}, M., {Kodama}, T., {Tadaki}, K.-i., {Koyama}, Y., \& {Tanaka}, I.
  2012, \apj, 757, 15

\bibitem[{{Henriques} {et~al.}(2012){Henriques}, {White}, {Lemson}, {Thomas},
  {Guo}, {Marleau}, \& {Overzier}}]{Henriques:2012}
{Henriques}, B.~M.~B., {White}, S.~D.~M., {Lemson}, G., {et~al.} 2012, \mnras,
  421, 2904

\bibitem[{{Henriques} {et~al.}(2015){Henriques}, {White}, {Thomas}, {Angulo},
  {Guo}, {Lemson}, {Springel}, \& {Overzier}}]{Henriques:2015}
{Henriques}, B.~M.~B., {White}, S.~D.~M., {Thomas}, P.~A., {et~al.} 2015,
  \mnras, 451, 2663

\bibitem[{{Johnson} {et~al.}(2007){Johnson}, {Almaini}, {Best}, \&
  {Dunlop}}]{Johnson:2007}
{Johnson}, O., {Almaini}, O., {Best}, P.~N., \& {Dunlop}, J. 2007, \mnras, 376,
  151

\bibitem[{{Kennicutt}(1998)}]{Kennicutt:1998}
{Kennicutt}, Jr., R.~C. 1998, \araa, 36, 189

\bibitem[{{Koekemoer} {et~al.}(2011){Koekemoer}, {Faber}, {Ferguson}, {Grogin},
  {Kocevski}, {Koo}, {Lai}, {Lotz}, {Lucas}, {McGrath}, {Ogaz}, {Rajan},
  {Riess}, {Rodney}, {Strolger}, {Casertano}, {Castellano}, {Dahlen},
  {Dickinson}, {Dolch}, {Fontana}, {Giavalisco}, {Grazian}, {Guo}, {Hathi},
  {Huang}, {van der Wel}, {Yan}, {Acquaviva}, {Alexander}, {Almaini}, {Ashby},
  {Barden}, {Bell}, {Bournaud}, {Brown}, {Caputi}, {Cassata}, {Challis},
  {Chary}, {Cheung}, {Cirasuolo}, {Conselice}, {Roshan Cooray}, {Croton},
  {Daddi}, {Dav{\'e}}, {de Mello}, {de Ravel}, {Dekel}, {Donley}, {Dunlop},
  {Dutton}, {Elbaz}, {Fazio}, {Filippenko}, {Finkelstein}, {Frazer}, {Gardner},
  {Garnavich}, {Gawiser}, {Gruetzbauch}, {Hartley}, {H{\"a}ussler},
  {Herrington}, {Hopkins}, {Huang}, {Jha}, {Johnson}, {Kartaltepe},
  {Khostovan}, {Kirshner}, {Lani}, {Lee}, {Li}, {Madau}, {McCarthy},
  {McIntosh}, {McLure}, {McPartland}, {Mobasher}, {Moreira}, {Mortlock},
  {Moustakas}, {Mozena}, {Nandra}, {Newman}, {Nielsen}, {Niemi}, {Noeske},
  {Papovich}, {Pentericci}, {Pope}, {Primack}, {Ravindranath}, {Reddy},
  {Renzini}, {Rix}, {Robaina}, {Rosario}, {Rosati}, {Salimbeni}, {Scarlata},
  {Siana}, {Simard}, {Smidt}, {Snyder}, {Somerville}, {Spinrad}, {Straughn},
  {Telford}, {Teplitz}, {Trump}, {Vargas}, {Villforth}, {Wagner}, {Wandro},
  {Wechsler}, {Weiner}, {Wiklind}, {Wild}, {Wilson}, {Wuyts}, \&
  {Yun}}]{Koekemoer:2011}
{Koekemoer}, A.~M., {Faber}, S.~M., {Ferguson}, H.~C., {et~al.} 2011, \apjs,
  197, 36

\bibitem[{{Koyama} {et~al.}(2013){Koyama}, {Kodama}, {Tadaki}, {Hayashi},
  {Tanaka}, {Smail}, {Tanaka}, \& {Kurk}}]{Koyama:2013}
{Koyama}, Y., {Kodama}, T., {Tadaki}, K.-i., {et~al.} 2013, \mnras, 428, 1551

\bibitem[{{Kravtsov} \& {Borgani}(2012)}]{Kravtsov:2012}
{Kravtsov}, A.~V., \& {Borgani}, S. 2012, \araa, 50, 353

\bibitem[{{Kriek} {et~al.}(2009){Kriek}, {van Dokkum}, {Labb{\'e}}, {Franx},
  {Illingworth}, {Marchesini}, \& {Quadri}}]{Kriek:2009a}
{Kriek}, M., {van Dokkum}, P.~G., {Labb{\'e}}, I., {et~al.} 2009, \apj, 700,
  221

\bibitem[{{Kubo} {et~al.}(2016){Kubo}, {Yamada}, {Ichikawa}, {Kajisawa},
  {Matsuda}, {Tanaka}, \& {Umehata}}]{Kubo:2016}
{Kubo}, M., {Yamada}, T., {Ichikawa}, T., {et~al.} 2016, \mnras, 455, 3333

\bibitem[{{Kurk} {et~al.}(2000){Kurk}, {R{\"o}ttgering}, {Pentericci}, {Miley},
  {van Breugel}, {Carilli}, {Ford}, {Heckman}, {McCarthy}, \&
  {Moorwood}}]{Kurk:2000}
{Kurk}, J.~D., {R{\"o}ttgering}, H.~J.~A., {Pentericci}, L., {et~al.} 2000,
  \aap, 358, L1

\bibitem[{{Larson} {et~al.}(1980){Larson}, {Tinsley}, \&
  {Caldwell}}]{Larson:1980}
{Larson}, R.~B., {Tinsley}, B.~M., \& {Caldwell}, C.~N. 1980, \apj, 237, 692

\bibitem[{{Leauthaud} {et~al.}(2010){Leauthaud}, {Finoguenov}, {Kneib},
  {Taylor}, {Massey}, {Rhodes}, {Ilbert}, {Bundy}, {Tinker}, {George}, {Capak},
  {Koekemoer}, {Johnston}, {Zhang}, {Cappelluti}, {Ellis}, {Elvis}, {Giodini},
  {Heymans}, {Le F{\`e}vre}, {Lilly}, {McCracken}, {Mellier},
  {R{\'e}fr{\'e}gier}, {Salvato}, {Scoville}, {Smoot}, {Tanaka}, {Van
  Waerbeke}, \& {Wolk}}]{Leauthaud:2010}
{Leauthaud}, A., {Finoguenov}, A., {Kneib}, J.-P., {et~al.} 2010, \apj, 709, 97

\bibitem[{{Lemaux} {et~al.}(2014){Lemaux}, {Cucciati}, {Tasca}, {Le F{\`e}vre},
  {Zamorani}, {Cassata}, {Garilli}, {Le Brun}, {Maccagni}, {Pentericci},
  {Thomas}, {Vanzella}, {Zucca}, {Amor{\'{\i}}n}, {Bardelli}, {Capak},
  {Cassar{\`a}}, {Castellano}, {Cimatti}, {Cuby}, {de la Torre}, {Durkalec},
  {Fontana}, {Giavalisco}, {Grazian}, {Hathi}, {Ilbert}, {Moreau}, {Paltani},
  {Ribeiro}, {Salvato}, {Schaerer}, {Scodeggio}, {Sommariva}, {Talia},
  {Taniguchi}, {Tresse}, {Vergani}, {Wang}, {Charlot}, {Contini}, {Fotopoulou},
  {Gal}, {Kocevski}, {L{\'o}pez-Sanjuan}, {Lubin}, {Mellier}, {Sadibekova}, \&
  {Scoville}}]{Lemaux:2014}
{Lemaux}, B.~C., {Cucciati}, O., {Tasca}, L.~A.~M., {et~al.} 2014, \aap, 572,
  A41

\bibitem[{{McCracken} {et~al.}(2012){McCracken}, {Milvang-Jensen}, {Dunlop},
  {Franx}, {Fynbo}, {Le F{\`e}vre}, {Holt}, {Caputi}, {Goranova}, {Buitrago},
  {Emerson}, {Freudling}, {Hudelot}, {L{\'o}pez-Sanjuan}, {Magnard}, {Mellier},
  {M{\o}ller}, {Nilsson}, {Sutherland}, {Tasca}, \& {Zabl}}]{McCracken:2012}
{McCracken}, H.~J., {Milvang-Jensen}, B., {Dunlop}, J., {et~al.} 2012, \aap,
  544, A156

\bibitem[{{McGee} {et~al.}(2009){McGee}, {Balogh}, {Bower}, {Font}, \&
  {McCarthy}}]{McGee:2009}
{McGee}, S.~L., {Balogh}, M.~L., {Bower}, R.~G., {Font}, A.~S., \& {McCarthy},
  I.~G. 2009, \mnras, 400, 937

\bibitem[{{McMullin} {et~al.}(2007){McMullin}, {Waters}, {Schiebel}, {Young},
  \& {Golap}}]{McMullin:2007}
{McMullin}, J.~P., {Waters}, B., {Schiebel}, D., {Young}, W., \& {Golap}, K.
  2007, in Astronomical Society of the Pacific Conference Series, Vol. 376,
  Astronomical Data Analysis Software and Systems XVI, ed. R.~A. {Shaw},
  F.~{Hill}, \& D.~J. {Bell}, 127

\bibitem[{{Mei} {et~al.}(2015){Mei}, {Scarlata}, {Pentericci}, {Newman},
  {Weiner}, {Ashby}, {Castellano}, {Conselice}, {Finkelstein}, {Galametz},
  {Grogin}, {Koekemoer}, {Huertas-Company}, {Lani}, {Lucas}, {Papovich},
  {Rafelski}, \& {Teplitz}}]{Mei:2015}
{Mei}, S., {Scarlata}, C., {Pentericci}, L., {et~al.} 2015, \apj, 804, 117

\bibitem[{{Miley} \& {De Breuck}(2008)}]{Miley:2008}
{Miley}, G., \& {De Breuck}, C. 2008, \aapr, 15, 67

\bibitem[{{Muldrew} {et~al.}(2015){Muldrew}, {Hatch}, \&
  {Cooke}}]{Muldrew:2015}
{Muldrew}, S.~I., {Hatch}, N.~A., \& {Cooke}, E.~A. 2015, \mnras, 452, 2528

\bibitem[{{Munari} {et~al.}(2013){Munari}, {Biviano}, {Borgani}, {Murante}, \&
  {Fabjan}}]{Munari:2013}
{Munari}, E., {Biviano}, A., {Borgani}, S., {Murante}, G., \& {Fabjan}, D.
  2013, \mnras, 430, 2638

\bibitem[{{Murray} {et~al.}(2013){Murray}, {Power}, \&
  {Robotham}}]{Murray:2013}
{Murray}, S.~G., {Power}, C., \& {Robotham}, A.~S.~G. 2013, Astronomy and
  Computing, 3, 23

\bibitem[{{Muzzin} {et~al.}(2013){Muzzin}, {Marchesini}, {Stefanon}, {Franx},
  {Milvang-Jensen}, {Dunlop}, {Fynbo}, {Brammer}, {Labb{\'e}}, \& {van
  Dokkum}}]{Muzzin:2013a}
{Muzzin}, A., {Marchesini}, D., {Stefanon}, M., {et~al.} 2013, \apjs, 206, 8

\bibitem[{{Navarro} {et~al.}(1997){Navarro}, {Frenk}, \&
  {White}}]{Navarro:1997}
{Navarro}, J.~F., {Frenk}, C.~S., \& {White}, S.~D.~M. 1997, \apj, 490, 493

\bibitem[{{Negrello} {et~al.}(2014){Negrello}, {Hopwood}, {Dye}, {Cunha},
  {Serjeant}, {Fritz}, {Rowlands}, {Fleuren}, {Bussmann}, {Cooray},
  {Dannerbauer}, {Gonzalez-Nuevo}, {Lapi}, {Omont}, {Amber}, {Auld}, {Baes},
  {Buttiglione}, {Cava}, {Danese}, {Dariush}, {De Zotti}, {Dunne}, {Eales},
  {Ibar}, {Ivison}, {Kim}, {Leeuw}, {Maddox}, {Micha{\l}owski}, {Massardi},
  {Pascale}, {Pohlen}, {Rigby}, {Smith}, {Sutherland}, {Temi}, \&
  {Wardlow}}]{Negrello:2014}
{Negrello}, M., {Hopwood}, R., {Dye}, S., {et~al.} 2014, \mnras, 440, 1999

\bibitem[{{Newman} {et~al.}(2014){Newman}, {Ellis}, {Andreon}, {Treu},
  {Raichoor}, \& {Trinchieri}}]{Newman:2014}
{Newman}, A.~B., {Ellis}, R.~S., {Andreon}, S., {et~al.} 2014, \apj, 788, 51

\bibitem[{{Overzier} {et~al.}(2005){Overzier}, {Harris}, {Carilli},
  {Pentericci}, {R{\"o}ttgering}, \& {Miley}}]{Overzier:2005}
{Overzier}, R.~A., {Harris}, D.~E., {Carilli}, C.~L., {et~al.} 2005, \aap, 433,
  87

\bibitem[{{Papovich} {et~al.}(2010){Papovich}, {Momcheva}, {Willmer},
  {Finkelstein}, {Finkelstein}, {Tran}, {Brodwin}, {Dunlop}, {Farrah}, {Khan},
  {Lotz}, {McCarthy}, {McLure}, {Rieke}, {Rudnick}, {Sivanandam}, {Pacaud}, \&
  {Pierre}}]{Papovich:2010}
{Papovich}, C., {Momcheva}, I., {Willmer}, C.~N.~A., {et~al.} 2010, \apj, 716,
  1503

\bibitem[{{Papovich} {et~al.}(2012){Papovich}, {Bassett}, {Lotz}, {van der
  Wel}, {Tran}, {Finkelstein}, {Bell}, {Conselice}, {Dekel}, {Dunlop}, {Guo},
  {Faber}, {Farrah}, {Ferguson}, {Finkelstein}, {H{\"a}ussler}, {Kocevski},
  {Koekemoer}, {Koo}, {McGrath}, {McLure}, {McIntosh}, {Momcheva}, {Newman},
  {Rudnick}, {Weiner}, {Willmer}, \& {Wuyts}}]{Papovich:2012}
{Papovich}, C., {Bassett}, R., {Lotz}, J.~M., {et~al.} 2012, \apj, 750, 93

\bibitem[{{Peng} {et~al.}(2010){Peng}, {Ho}, {Impey}, \& {Rix}}]{PengC:2010}
{Peng}, C.~Y., {Ho}, L.~C., {Impey}, C.~D., \& {Rix}, H.-W. 2010, \aj, 139,
  2097

\bibitem[{{Planck Collaboration} {et~al.}(2015){Planck Collaboration}, {Ade},
  {Aghanim}, {Arnaud}, {Ashdown}, {Aumont}, {Baccigalupi}, {Banday},
  {Barreiro}, {Bartlett}, \& et~al.}]{PlanckXIII:2015}
{Planck Collaboration}, {Ade}, P.~A.~R., {Aghanim}, N., {et~al.} 2015, ArXiv
  e-prints, arXiv:1502.01589

\bibitem[{{Ranalli} {et~al.}(2003){Ranalli}, {Comastri}, \&
  {Setti}}]{Ranalli:2003}
{Ranalli}, P., {Comastri}, A., \& {Setti}, G. 2003, \aap, 399, 39

\bibitem[{{Reichert} {et~al.}(2011){Reichert}, {B{\"o}hringer}, {Fassbender},
  \& {M{\"u}hlegger}}]{Reichert:2011}
{Reichert}, A., {B{\"o}hringer}, H., {Fassbender}, R., \& {M{\"u}hlegger}, M.
  2011, \aap, 535, A4

\bibitem[{{Rodighiero} {et~al.}(2011){Rodighiero}, {Daddi}, {Baronchelli},
  {Cimatti}, {Renzini}, {Aussel}, {Popesso}, {Lutz}, {Andreani}, {Berta},
  {Cava}, {Elbaz}, {Feltre}, {Fontana}, {F{\"o}rster Schreiber},
  {Franceschini}, {Genzel}, {Grazian}, {Gruppioni}, {Ilbert}, {Le Floch},
  {Magdis}, {Magliocchetti}, {Magnelli}, {Maiolino}, {McCracken}, {Nordon},
  {Poglitsch}, {Santini}, {Pozzi}, {Riguccini}, {Tacconi}, {Wuyts}, \&
  {Zamorani}}]{Rodighiero:2011}
{Rodighiero}, G., {Daddi}, E., {Baronchelli}, I., {et~al.} 2011, \apjl, 739,
  L40

\bibitem[{{Ruel} {et~al.}(2014){Ruel}, {Bazin}, {Bayliss}, {Brodwin}, {Foley},
  {Stalder}, {Aird}, {Armstrong}, {Ashby}, {Bautz}, {Benson}, {Bleem},
  {Bocquet}, {Carlstrom}, {Chang}, {Chapman}, {Cho}, {Clocchiatti}, {Crawford},
  {Crites}, {de Haan}, {Desai}, {Dobbs}, {Dudley}, {Forman}, {George},
  {Gladders}, {Gonzalez}, {Halverson}, {Harrington}, {High}, {Holder},
  {Holzapfel}, {Hrubes}, {Jones}, {Joy}, {Keisler}, {Knox}, {Lee}, {Leitch},
  {Liu}, {Lueker}, {Luong-Van}, {Mantz}, {Marrone}, {McDonald}, {McMahon},
  {Mehl}, {Meyer}, {Mocanu}, {Mohr}, {Montroy}, {Murray}, {Natoli},
  {Nurgaliev}, {Padin}, {Plagge}, {Pryke}, {Reichardt}, {Rest}, {Ruhl},
  {Saliwanchik}, {Saro}, {Sayre}, {Schaffer}, {Shaw}, {Shirokoff}, {Song}, {{\v
  S}uhada}, {Spieler}, {Stanford}, {Staniszewski}, {Starsk}, {Story}, {Stubbs},
  {van Engelen}, {Vanderlinde}, {Vieira}, {Vikhlinin}, {Williamson}, {Zahn}, \&
  {Zenteno}}]{Ruel:2014}
{Ruel}, J., {Bazin}, G., {Bayliss}, M., {et~al.} 2014, \apj, 792, 45

\bibitem[{{Salpeter}(1955)}]{Salpeter:1955}
{Salpeter}, E.~E. 1955, \apj, 121, 161

\bibitem[{{Santos} {et~al.}(2015){Santos}, {Altieri}, {Valtchanov}, {Nastasi},
  {B{\"o}hringer}, {Cresci}, {Elbaz}, {Fassbender}, {Rosati}, {Tozzi}, \&
  {Verdugo}}]{Santos:2015}
{Santos}, J.~S., {Altieri}, B., {Valtchanov}, I., {et~al.} 2015, \mnras, 447,
  L65

\bibitem[{{Saro} {et~al.}(2013){Saro}, {Mohr}, {Bazin}, \& {Dolag}}]{Saro:2013}
{Saro}, A., {Mohr}, J.~J., {Bazin}, G., \& {Dolag}, K. 2013, \apj, 772, 47

\bibitem[{{Schreiber} {et~al.}(2015){Schreiber}, {Pannella}, {Elbaz},
  {B{\'e}thermin}, {Inami}, {Dickinson}, {Magnelli}, {Wang}, {Aussel}, {Daddi},
  {Juneau}, {Shu}, {Sargent}, {Buat}, {Faber}, {Ferguson}, {Giavalisco},
  {Koekemoer}, {Magdis}, {Morrison}, {Papovich}, {Santini}, \&
  {Scott}}]{Schreiber:2015}
{Schreiber}, C., {Pannella}, M., {Elbaz}, D., {et~al.} 2015, \aap, 575, A74

\bibitem[{{Sersic}(1968)}]{Sersic:1968}
{Sersic}, J.~L. 1968, {Atlas de galaxias australes}

\bibitem[{{Sharples} {et~al.}(2013){Sharples}, {Bender}, {Agudo Berbel},
  {Bezawada}, {Castillo}, {Cirasuolo}, {Davidson}, {Davies}, {Dubbeldam},
  {Fairley}, {Finger}, {F{\"o}rster Schreiber}, {Gonte}, {Hess}, {Jung},
  {Lewis}, {Lizon}, {Muschielok}, {Pasquini}, {Pirard}, {Popovic}, {Ramsay},
  {Rees}, {Richter}, {Riquelme}, {Rodrigues}, {Saviane}, {Schlichter},
  {Schmidtobreick}, {Segovia}, {Smette}, {Szeifert}, {van Kesteren}, {Wegner},
  \& {Wiezorrek}}]{Sharples:2013}
{Sharples}, R., {Bender}, R., {Agudo Berbel}, A., {et~al.} 2013, The Messenger,
  151, 21

\bibitem[{{Sharples} {et~al.}(2004){Sharples}, {Bender}, {Lehnert}, {Ramsay
  Howat}, {Bremer}, {Davies}, {Genzel}, {Hofmann}, {Ivison}, {Saglia}, \&
  {Thatte}}]{Sharples:2004}
{Sharples}, R.~M., {Bender}, R., {Lehnert}, M.~D., {et~al.} 2004, in \procspie,
  Vol. 5492, Ground-based Instrumentation for Astronomy, ed. A.~F.~M.
  {Moorwood} \& M.~{Iye}, 1179--1186

\bibitem[{{Smol{\v c}i{\'c}} {et~al.}(2014){Smol{\v c}i{\'c}}, {Ciliegi},
  {Jeli{\'c}}, {Bondi}, {Schinnerer}, {Carilli}, {Riechers}, {Salvato},
  {Brkovi{\'c}}, {Capak}, {Ilbert}, {Karim}, {McCracken}, \&
  {Scoville}}]{Smolvcic:2014}
{Smol{\v c}i{\'c}}, V., {Ciliegi}, P., {Jeli{\'c}}, V., {et~al.} 2014, \mnras,
  443, 2590

\bibitem[{{Spitler} {et~al.}(2012){Spitler}, {Labb{\'e}}, {Glazebrook},
  {Persson}, {Monson}, {Papovich}, {Tran}, {Poole}, {Quadri}, {van Dokkum},
  {Kelson}, {Kacprzak}, {McCarthy}, {Murphy}, {Straatman}, \&
  {Tilvi}}]{Spitler:2012}
{Spitler}, L.~R., {Labb{\'e}}, I., {Glazebrook}, K., {et~al.} 2012, \apjl, 748,
  L21

\bibitem[{{Springel} {et~al.}(2005){Springel}, {Di Matteo}, \&
  {Hernquist}}]{Springel:2005}
{Springel}, V., {Di Matteo}, T., \& {Hernquist}, L. 2005, \apjl, 620, L79

\bibitem[{{Stanford} {et~al.}(2012){Stanford}, {Brodwin}, {Gonzalez},
  {Zeimann}, {Stern}, {Dey}, {Eisenhardt}, {Snyder}, \&
  {Mancone}}]{Stanford:2012}
{Stanford}, S.~A., {Brodwin}, M., {Gonzalez}, A.~H., {et~al.} 2012, \apj, 753,
  164

\bibitem[{{Steidel} {et~al.}(1998){Steidel}, {Adelberger}, {Dickinson},
  {Giavalisco}, {Pettini}, \& {Kellogg}}]{Steidel:1998}
{Steidel}, C.~C., {Adelberger}, K.~L., {Dickinson}, M., {et~al.} 1998, \apj,
  492, 428

\bibitem[{{Strazzullo} {et~al.}(2013){Strazzullo}, {Gobat}, {Daddi}, {Onodera},
  {Carollo}, {Dickinson}, {Renzini}, {Arimoto}, {Cimatti}, {Finoguenov}, \&
  {Chary}}]{Strazzullo:2013}
{Strazzullo}, V., {Gobat}, R., {Daddi}, E., {et~al.} 2013, \apj, 772, 118

\bibitem[{{Tanaka} {et~al.}(2011){Tanaka}, {Breuck}, {Kurk}, {Taniguchi},
  {Kodama}, {Matsuda}, {Packham}, {Zirm}, {Kajisawa}, {Ichikawa}, {Seymour},
  {Stern}, {Stockton}, {Venemans}, \& {Vernet}}]{Tanaka:2011}
{Tanaka}, I., {Breuck}, C.~D., {Kurk}, J.~D., {et~al.} 2011, \pasj, 63, 415

\bibitem[{{Taniguchi} {et~al.}(2015){Taniguchi}, {Kajisawa}, {Kobayashi},
  {Shioya}, {Nagao}, {Capak}, {Aussel}, {Ichikawa}, {Murayama}, {Scoville},
  {Ilbert}, {Salvato}, {Sanders}, {Mobasher}, {Miyazaki}, {Komiyama}, {Le
  F{\`e}vre}, {Tasca}, {Lilly}, {Carollo}, {Renzini}, {Rich}, {Schinnerer},
  {Kaifu}, {Karoji}, {Arimoto}, {Okamura}, {Ohta}, {Shimasaku}, \&
  {Hayashino}}]{Taniguchi:2015}
{Taniguchi}, Y., {Kajisawa}, M., {Kobayashi}, M.~A.~R., {et~al.} 2015, \pasj,
  67, 104

\bibitem[{{Thomas} {et~al.}(2005){Thomas}, {Maraston}, {Bender}, \& {Mendes de
  Oliveira}}]{Thomas:2005}
{Thomas}, D., {Maraston}, C., {Bender}, R., \& {Mendes de Oliveira}, C. 2005,
  \apj, 621, 673

\bibitem[{{Tran} {et~al.}(2010){Tran}, {Papovich}, {Saintonge}, {Brodwin},
  {Dunlop}, {Farrah}, {Finkelstein}, {Finkelstein}, {Lotz}, {McLure},
  {Momcheva}, \& {Willmer}}]{Tran:2010}
{Tran}, K.-V.~H., {Papovich}, C., {Saintonge}, A., {et~al.} 2010, \apjl, 719,
  L126

\bibitem[{{Trenti} {et~al.}(2012){Trenti}, {Bradley}, {Stiavelli}, {Shull},
  {Oesch}, {Bouwens}, {Mu{\~n}oz}, {Romano-Diaz}, {Treu}, {Shlosman}, \&
  {Carollo}}]{Trenti:2012}
{Trenti}, M., {Bradley}, L.~D., {Stiavelli}, M., {et~al.} 2012, \apj, 746, 55

\bibitem[{{Uchimoto} {et~al.}(2012){Uchimoto}, {Yamada}, {Kajisawa}, {Kubo},
  {Ichikawa}, {Matsuda}, {Akiyama}, {Hayashino}, {Konishi}, {Nishimura},
  {Omata}, {Suzuki}, {Tanaka}, {Tokoku}, \& {Yoshikawa}}]{Uchimoto:2012}
{Uchimoto}, Y.~K., {Yamada}, T., {Kajisawa}, M., {et~al.} 2012, \apj, 750, 116

\bibitem[{{Valentino} {et~al.}(2015){Valentino}, {Daddi}, {Strazzullo},
  {Gobat}, {Onodera}, {Bournaud}, {Juneau}, {Renzini}, {Arimoto}, {Carollo}, \&
  {Zanella}}]{Valentino:2015}
{Valentino}, F., {Daddi}, E., {Strazzullo}, V., {et~al.} 2015, \apj, 801, 132

\bibitem[{{van der Burg} {et~al.}(2014){van der Burg}, {Muzzin}, {Hoekstra},
  {Wilson}, {Lidman}, \& {Yee}}]{vanderBurg:2014}
{van der Burg}, R.~F.~J., {Muzzin}, A., {Hoekstra}, H., {et~al.} 2014, \aap,
  561, A79

\bibitem[{{van der Burg} {et~al.}(2013){van der Burg}, {Muzzin}, {Hoekstra},
  {Lidman}, {Rettura}, {Wilson}, {Yee}, {Hildebrandt}, {Marchesini},
  {Stefanon}, {Demarco}, \& {Kuijken}}]{vanderBurg:2013}
---. 2013, \aap, 557, A15

\bibitem[{{van der Wel} {et~al.}(2014){van der Wel}, {Franx}, {van Dokkum},
  {Skelton}, {Momcheva}, {Whitaker}, {Brammer}, {Bell}, {Rix}, {Wuyts},
  {Ferguson}, {Holden}, {Barro}, {Koekemoer}, {Chang}, {McGrath},
  {H{\"a}ussler}, {Dekel}, {Behroozi}, {Fumagalli}, {Leja}, {Lundgren},
  {Maseda}, {Nelson}, {Wake}, {Patel}, {Labb{\'e}}, {Faber}, {Grogin}, \&
  {Kocevski}}]{vanderWel:2014}
{van der Wel}, A., {Franx}, M., {van Dokkum}, P.~G., {et~al.} 2014, \apj, 788,
  28

\bibitem[{{van Dokkum} {et~al.}(2003){van Dokkum}, {F{\"o}rster Schreiber},
  {Franx}, {Daddi}, {Illingworth}, {Labb{\'e}}, {Moorwood}, {Rix},
  {R{\"o}ttgering}, {Rudnick}, {van der Wel}, {van der Werf}, \& {van
  Starkenburg}}]{vanDokkum:2003}
{van Dokkum}, P.~G., {F{\"o}rster Schreiber}, N.~M., {Franx}, M., {et~al.}
  2003, \apjl, 587, L83

\bibitem[{{Venemans} {et~al.}(2007){Venemans}, {R{\"o}ttgering}, {Miley}, {van
  Breugel}, {de Breuck}, {Kurk}, {Pentericci}, {Stanford}, {Overzier}, {Croft},
  \& {Ford}}]{Venemans:2007}
{Venemans}, B.~P., {R{\"o}ttgering}, H.~J.~A., {Miley}, G.~K., {et~al.} 2007,
  \aap, 461, 823

\bibitem[{{Webb} {et~al.}(2015){Webb}, {Noble}, {DeGroot}, {Wilson}, {Muzzin},
  {Bonaventura}, {Cooper}, {Delahaye}, {Foltz}, {Lidman}, {Surace}, {Yee},
  {Chapman}, {Dunne}, {Geach}, {Hayden}, {Hildebrandt}, {Huang}, {Pope},
  {Smith}, {Perlmutter}, \& {Tudorica}}]{Webb:2015}
{Webb}, T., {Noble}, A., {DeGroot}, A., {et~al.} 2015, \apj, 809, 173

\bibitem[{{Wegner} \& {Muschielok}(2008)}]{Wegner:2008}
{Wegner}, M., \& {Muschielok}, B. 2008, in \procspie, Vol. 7019, Advanced
  Software and Control for Astronomy II, 70190T

\bibitem[{{White} {et~al.}(2010){White}, {Cohn}, \& {Smit}}]{White:2010}
{White}, M., {Cohn}, J.~D., \& {Smit}, R. 2010, \mnras, 408, 1818

\bibitem[{{Williams} {et~al.}(2009){Williams}, {Quadri}, {Franx}, {van Dokkum},
  \& {Labb{\'e}}}]{Williams:2009}
{Williams}, R.~J., {Quadri}, R.~F., {Franx}, M., {van Dokkum}, P., \&
  {Labb{\'e}}, I. 2009, \apj, 691, 1879

\bibitem[{{Yuan} {et~al.}(2014){Yuan}, {Nanayakkara}, {Kacprzak}, {Tran},
  {Glazebrook}, {Kewley}, {Spitler}, {Poole}, {Labb{\'e}}, {Straatman}, \&
  {Tomczak}}]{YuanT:2014}
{Yuan}, T., {Nanayakkara}, T., {Kacprzak}, G.~G., {et~al.} 2014, \apjl, 795,
  L20

\end{thebibliography}

\end{document}